\documentclass{article}
\usepackage[utf8]{inputenc}
\usepackage[english]{babel}

\usepackage[letterpaper,top=2cm,bottom=2cm,left=3cm,right=3cm,marginparwidth=1.75cm]{geometry}
 
\usepackage{lmodern,textcomp}
\usepackage{amsmath,amsfonts }
\usepackage[table,xcdraw,dvipsnames]{xcolor}
\usepackage[colorlinks=true, allcolors=blue]{hyperref}
\usepackage{natbib}
\usepackage{csquotes}
\usepackage{graphicx}
\usepackage{caption}
\usepackage{subcaption}
\usepackage{pdflscape}

\newcommand\blfootnote[1]{%
  \begingroup
  \renewcommand\thefootnote{}\footnote{#1}%
  \addtocounter{footnote}{-1}%
  \endgroup
}

\newcommand{\ca}{$^{\dagger}$}

\newcommand{\toMILA}{$^1$}
\newcommand{\toATI}{$^2$}
\newcommand{\toUdeM}{$^{3}$}
\newcommand{\toUOxford}{$^{4}$}
\newcommand{\toCIFAR}{$^{5}$}
\newcommand{\toINET}{$^{6}$}

\title{(Private)-Retroactive Carbon Pricing [(P)ReCaP]:\\ A Market-based Approach for Climate Finance and Risk Assessment}

\author{
Yoshua Bengio\toMILA$^,$\toUdeM$^,$\toCIFAR,
Prateek Gupta\toMILA$^,$\toATI$^,$\toUOxford, 
Dylan Radovic \toUOxford, 
Maarten Scholl \toUOxford$^,$\toINET, \\
Andrew Williams\toMILA$^,$\toUdeM,
Christian Schroeder de Witt \toUOxford$^,$\ca,
Tianyu Zhang \toMILA$^,$\toUdeM$^,$\ca, 
Yang Zhang\toMILA
}

\begin{document}

\maketitle

\blfootnote{
authors are listed in alphabetical order of their last names, \\ 
\toMILA Mila - Québec AI Institute,
\toATI The Alan Turing Institute,
\toUdeM Universit\'e de Montr\'eal,
\toUOxford University of Oxford,
\toCIFAR CIFAR Fellow
\toINET Institute for New Economic Thinking, Oxford Martin School
.
\\ 
\ca Corresponding authors \texttt{schroederdewitt@gmail.com},  \texttt{tianyu.zhang@mila.quebec}. 
}

\begin{abstract} 
    Insufficient Social Cost of Carbon (SCC) estimation methods and short-term decision-making horizons have hindered the ability of carbon emitters to properly correct for the negative externalities of climate change, as well as the capacity of nations to balance economic and climate policy.
    To overcome these limitations, we introduce Retrospective Social Cost of Carbon Updating (ReSCCU), a novel mechanism that corrects for these limitations as empirically measured evidence is collected. 
    To implement ReSCCU in the context of carbon taxation, we propose Retroactive Carbon Pricing (ReCaP), a market mechanism in which polluters offload the payment of ReSCCU adjustments to insurers.
    To alleviate systematic risks and minimize government involvement, we introduce the Private ReCaP (PReCaP) prediction market, which could see real-world implementation based on the engagement of a few high net-worth individuals or independent institutions.
\end{abstract}

\tableofcontents
\section{Executive Summary}

As is evident through numerous studies, human civilization faces a severe threat and amplification of potentially extinction-level threats due to the harmful effects of climate change. 
As a result, individuals and communities worldwide have begun to change their outlook and objectives to minimize their impact on the environment.
At the same time, nations are aiming to decouple their GDP growth from carbon emissions through investment in climate mitigation and adaptation projects.
However, due to inadequate Social Cost of Carbon (SCC) estimation methods, i.e., long-term climate impacts of emissions and shorter planning horizons of corporations and governments, it has been challenging to accurately attribute negative climate impacts to the polluters. 

It is widely acknowledged that the estimation of the SCC is inherently laden with uncertainties, and numerous works have discussed the implications of the fact that we can not fully ascertain the cost of future climate damages. In this paper, we add several novel perspectives to this debate. 
First of all, we show that fundamental limits of predictability do not limit existing methods of predicting the SCC. Rather, as of today, SCC modeling is limited by a lack of incentives for model innovation and data access obstacles that prevent current models from being calibrated to past climate damages.

Secondly, we explore the implications of the fact that the stakeholders like corporations, governments, and households operate on a shorter planning horizon (typically, 2-5 years) compared to the harmful impacts of carbon emissions, which are evident only decades later. This \textit{Tragedy of the Horizon} has fundamental implications on existing approaches to SCC-based decision-making. 

We show that our novel Retrospective Social Cost of Carbon Updating (ReSCCU) mechanism fundamentally addresses the above issues. We propose that the mechanism supersedes the SCC to ensure that every dollar paid by the polluter for carbon emissions is attributed to climate harm. 

We further propose Retroactive Carbon Pricing (ReCaP), a government-enforced carbon taxation scheme implementing ReSCCU that enables polluters of any market size, to participate, and sell the payment of yearly ReSCCU adjustments to insurers, who swap payments with the government.
ReCaP invites large financial institutes (e.g., investment banks and insurance companies) to insure the carbon credits listed by the suppliers and take on the responsibility of future ReSCCU adjustments between the government and itself. \footnote{Note that these aren't strictly financial insurances as they do not actually ``insure" any real-world assets in the conventional sense.}
Here the government acts as an exchange, and the competition among the financial institutes fosters the model innovation while maintaining the polluters-pay principle. 

Finally, to circumvent the risk of political influence affecting ReCaP, we propose Private ReCaP (PReCaP) in the context of the private markets. 
PReCaP sets up a private RetroExchange that enables polluters to buy insured carbon credits from insurers and the carbon credit supplier through a three-agent auction mechanism. 
The resulting approach requires voluntary participation and fits well with the ideals of voluntary carbon markets (VCM).

We hope to stimulate discussion among policy designers, computer scientists, economists, and climate scientists to consider an alternative perspective on SCC and carbon taxation through this work.
We also put forth a call to action from governments and investors to set up PReCaP that can help bring in a critical mass of polluters, credit suppliers and financial institutes to achieve the momentum required for its feasible operations. 

\section{Introduction}

Policymakers have convened~\citep{UNFCCC2015} to discuss how to limit the increase in global mean temperature to 2°C or less by decreasing greenhouse gases (GHG).
Confining global warming to these safe levels by the end of the century requires unprecedented mitigation and adaptation efforts. 
Such efforts include, but are not limited to
\begin{itemize}
    \item decarbonization in the global energy system;
    \item novel low-carbon technologies in all industry sectors;
    \item rectification of individual consumption patterns;
    \item financial adaptation transfers to vulnerable communities.
\end{itemize}

Unfortunately, the current incentive structures in social-economic and political settings are largely misaligned with what is needed to reach the desired level of emissions while promoting well-being~\citep{OECD2019}. 
This outcome arises from two key challenges.

From a theoretical perspective, free markets’ inability to maximize society’s welfare deems climate change a market failure~\citep{Benjamin2007}. 
For global market participants, climate change is primarily an \textit{externality}, signifying that those who emit greenhouse gases are not directly liable for the resulting damage.
Since rising GHG emissions are considered external to markets, failure to internalize marginal social damages often leads to the overproduction of such negative externalities. 
Economists have suggested setting a price on GHG emissions to alter incentives and shift capital towards low-carbon opportunities.
Various pricing mechanisms have been introduced to internalize climate damage into market economies (see Appendix A for a detailed discussion of existing approaches). 

More importantly, from a practical perspective, any climate change mitigation or  adaptation effort requires the implementation of long-term measures by decision-makers at all levels of governance.
However, corporate and political cycles driven by executives and politicians' decisions do not normally look more than the horizon beyond their responsible mandate period. 
This horizon varies in general from two to five years in the future.

The apparent tension between real-world decision-making under short planning horizons and the long-term challenges posed by climate change has been termed as \textit{Tragedy of the Horizon}~\citep{Carney2015}.
An obvious solution to overcome this tragedy, although non-trivial, requires \textit{bringing the future into the present}~\citep{Carney2015}, meaning emphasizing the importance of today's decision-making on future catastrophic impacts of climate change through implementing the \textit{polluter pays principle (PPP)}.
Doing so in practice helps establish an international consensus on how to align polluter incentives with safe planetary pathways.

An estimator for lifetime emission damages is the \textbf{social cost of carbon (SCC)}, a concept introduced under the Obama administration.  
In the US, the 2009 Executive Order 12866 requires government agencies to consider the costs and benefits of any potential regulations. 
Many factors are difficult to internalize and estimate in monetary values, so only those regulations whose benefits would justify the cost should be proposed.
As a result, President Obama established the SCC as the fundamental assessor of the cost-benefit analysis of such policy evaluations.
Since then, several other nations and financial institutes like impact investment funds have been using an estimate of the SCC for risk assessment and financing activities. 

More formally, the SCC represents the present value of all future damages (in perpetuity) incurred by an incremental ton of carbon-equivalent emissions.
Thus, it sets a price on one tonne of carbon-equivalent emission that the emitter should pay to appropriately account for the environmental-specific negative externalities (or indirect environmental damages). 
However, there are several difficulties in enforcing this scheme, from insufficient scientific modeling to estimate damages and political hurdles to future uncertainties regarding technological development. 

To this end, we propose novel market mechanisms to enforce the PPP while fostering accurate estimation of the SCC, thereby fairly internalizing the harmful environmental damages. 
The rest of the paper is organized as follows.
In section~\ref{sec:scc-insuff}, we argue that there are inherent and severe difficulties in estimating the SCC due to the political risks and modeling uncertainties. 
To overcome the unpredictable nature of SCC, we propose the Retrospective Social Cost of Carbon Updating (ReSCCU) mechanism in Section~\ref{sec:resccu}.
We argue, in section~\ref{sec:resccu-tax}, that ReSCCU is a form of a carbon tax, following which we introduce Retroactive Carbon Pricing (ReCaP) conceptually in the context of PPP-conformant carbon taxation (via ReSCCU). 
Further, we translate ReCaP into a private market setting in section~\ref{sec:precap} on Private-ReCaP (PReCaP), and show how this setting overcomes the limits of risk mitigation as well as political and implementation hurdles, thereby giving rise to improved, market-driven SCC estimates and stimulating model innovation.
Finally, we conclude in Section~\ref{sec:conclusion} suggesting that as our private market setting does not necessitate government involvement, PReCaP constitutes a feasible first step toward pricing carbon while overcoming the Tragedy of Horizon.

\section{Insufficient signals to price the SCC}
\label{sec:scc-insuff}

While the SCC is inherently designed to bring future damages into the present, it also faces several challenges and limitations in practice.

\subsection{Lack of consensus among nations}

First of all, determining the SCC requires modeling the dynamics of the physical climate and socioeconomic pathways. 
While the transient climate response to added emissions is generally well understood, the family of socioeconomic models underlying SCC estimation is known to be deficient. 

Ethical and political choices surrounding whose damages to include introduce greater error in SCC estimates, which are harder to mitigate.
It is also challenging to determine the discount rate to apply to these expected damages.
There is a great degree of heterogeneity in different countries' practices. 
For example, successive US governments have triggered substantial fluctuations in \textit{social discounting rates}. 
A social discounting rate is a number (ranging between zero and one) that weighs the importance of costs occurring in the future - a choice that usually reflects considerations of ethical values. 
There is also perpetual disagreement on whether damages outside the US should be considered. 
The resulting SCC prices determined by the US government have been largely erratic and are detached from what the Intergovernmental Panel on Climate Change (IPCC) would consider sufficiently high in order to constrain global warming to safe levels~\citep{ipcc_global_2018}.

Few countries are currently following the standard route of internalizing carbon externalities through taxation~\cite{Pigou1920}. 
Following the \textit{polluter pays principle}, carbon taxes should be set to cover the SCC fully. 
However, actual carbon tax rates implemented by countries such as Canada tend to be at significantly lower levels that cover only a fraction of the SCC~\citep{urban_canadian_2021}.

Another significant share of the world economy, including the UK and EU, has decided not to use the SCC in their mitigation efforts.
Instead, they let carbon markets determine the carbon price under a fixed external carbon budget. 
The reasons for eschewing the SCC are partially due to the difficulty of establishing it quantitatively and, ultimately, countries' uneasiness to rely on a single number when assessing respective and collective climate risks (see, for example, the UK's debate on the \textit{shadow cost of carbon} \citep{economics_group_social_2007}). 
Such target-consistent approaches to carbon prices aim to ensure that cumulative emissions remain in accordance with safe mitigation pathways, but do not, even in theory, make polluters directly liable for the associated damages. 
In practice, carbon markets, such as the EU ETS, have suffered from a variety of pricing inefficiencies, not least due to unanticipated interactions with green subsidies \citep{braathen_interactions_2011}. 

An emerging alternative to government-run carbon markets are \textit{voluntary carbon markets (VCMs)}. 
VCMs are privately-run markets that allow suppliers of carbon credits to find suitable buyers.
Such markets currently consist largely of online directories listing buy and sell orders. 
The \textit{Taskforce for Scaling Voluntary Carbon Markets (TSVCM)} is an international organization aiming at standardizing carbon credits into derivatives that can be traded on mainstream exchanges just as other commodities. 
While currently in the order of a few million dollars annually, voluntary carbon markets are estimated to reach trading volumes of several billion dollars a year by $2030$ \citep{tsvcm_taskforce_2021}. 
As other carbon markets, VCMs pricing is subject to demand and supply and not directly related to the underlying damages of emitting greenhouse gases. 

While new initiatives in mitigating the SCC continue to emerge in most developed economies, many polluting economies continue to be subject to few or no carbon damage liabilities.
Moral hazard problems further highlight the widespread political barriers associated with implementing international organization-driven carbon pricing in the first place.

\subsection{Lack of signals in the market}

Aside from direct carbon pricing mechanisms, we now discuss the limitations of a few existing long-term mechanisms that, albeit indirectly, may reveal signals relevant to bringing the future into the present for carbon pricing. 

First of all, financial markets offer tradable derivatives with long-term horizons. 
A critical financial benchmark for climate change are interest rate benchmarks, such as LIBOR/SOFR (EURIBOR/ESTR), which are used as reference point for lenders across the board.
While short-term by construction, such benchmark rates can be leveraged by long-term derivatives. 
For example, derivatives on EURIBOR rates, had contracts with maturities up to $50$ years, and other G5 currencies benchmarks have swap par rates available for maturities up to $30$ years.
Under the efficient market hypothesis, such derivatives should be priced with anticipated climate damage liabilities, particularly as climate change mitigation and adaptation necessitate long-term investments at unprecedented volumes.
However, few market participants actively trade such long-term derivatives, resulting in illiquidity that further undermines pricing accuracy.

Secondly, insurance companies have been increasingly challenged on several fronts in meeting need to accurately and timely factor climate-related risks into long-term insurance policies.
Insurance premiums need to price in climate risks to the underlying assets, thus, in principle, revealing an estimate for expected climate damages. 
Fundamentally, climate insurance risk is hard to diversify as catastrophic risks are naturally correlated across regions and asset classes. 

In addition, the insurance industry suffers from a general lack of model innovation and the models in use are usually drawn from the families of traditional models underlying SCC estimates. 
The potential model misspecification and limitation across the industry could correlate insurer default risks and further complicate reinsurance, thereby yielding systematic vulnerability.
In practice, insurance coverage greatly varies both geographically and across sectors, and the associated provider landscape is highly fragmented, making it difficult to extract unbiased price signals into the future. 

Last but not least, recent years have seen a surge of climate litigation cases brought to court. 
While presently sparse and moderately successful, such cases do increasingly include adaptation costs to future damages, thus shining the spotlight on select long-term risks. 
If such trials become commonplace, a standardized market for litigation insurance may arise that yields some long-term pricing signals. 
Whether this will happen in time, however, remains subject to speculation.\\

As illustrated above, while some progress in internalizing catastrophic impacts of climate change through carbon pricing has been made, existing approaches either do not explicitly account for actual climate damages over the long run, or suffer from political risks and inherent modeling difficulties. 

While some encouraging progress has been made in internalizing the SCC to some extent, public and private policies in general remain challenged facing the ``Tragedy of the Horizon", and do not adequately bring the consideration of future problems into today's decision-making. 

\section{Limitations on Carbon Tax and the SCC}

\label{sec:resccu-tax}


A \textit{carbon tax} is a mechanism through which a government charges polluters, usually companies, according to their GHG emissions. 
Carbon taxes aim to align \textit{private firms} with \textit{social good}.
First, charging a carbon tax pools funds that the government can use in order to pay for communal damages created due to pollution events.
It is established practice in international environmental law that a polluter should pay for the totality of the damages it is responsible for, as encoded in the PPP \citep{tokuc_rio_2013}. 

Second, the act of charging polluters may by itself deter polluters from \textit{overproduction} of goods, thus restoring market equilibria distorted by a lack of internalization of externalities.
Both of these aspects are captured by the concept of \textit{Pigouvian tax}~\citep{Pigou1920}, which internalizes externalities, such as pollution, by modifying the maximum of the consumer utility function. 
~\citet{Pigou1920} showed that the consumer's choice coincides with the socially optimal investment level if and only if Pigouvian tax is set to the marginal pollution damage. 

As we establish in the following section (Section~\ref{sec:resccu}), calculating the damage inflicted due to an additional ton of CO2 (or CO2-equivalent) released into the atmosphere is inherently difficult.
If a ton of CO2 is released into the atmosphere today, half of it will still remain thereafter $30$ years, and a fifth of it even after hundreds of years. 
Even if a polluter applies artificial technological means to remove an extra ton of CO2 from the atmosphere at a later date, it is possible that damage continues to occur due to system hysteresis. 
GHG pollution events are therefore inherently \textit{temporally extended over a long time horizon}. 
Pigouvian taxation was proposed long before anthropogenic climate change became a wide economic cause for concern. 
Consequently,~\citet{Pigou1920} did not consider the case that Pigouvian tax could be hard, or indeed impossible, to calculate and, furthermore, only be estimated under not-so-significant uncertainty. 

The question of whether to price in uncertainties in estimating the SCC to fix prices for Pigouvian tax has been raised soon after its adoption in 2009.
Indeed,~\citep{metcalf_design_2009} suggested that an optimally designed carbon tax would respond to uncertainty by continuously updating the tax rate ``as new information becomes available about the costs and benefits of reducing emissions''. 
In the face of uncertainty about the costs of abatement, the best available option was to ``utilize a crude estimate of the optimal rate and adjust the rate as new information arises''.
However,~\citet{metcalf_design_2009} does not distinguish between irreducible (aleatoric) prediction errors and epistemic errors, the latter being in principle avoidable in the limit of sufficient data collection and model calibration.

Further,~\citet{kousky_risk_2011} argue that SCC estimates should carry a \textit{risk premium}, reflecting that each abated ton of GHG emissions would also thin the fat-tail of low-probability, high-impact, catastrophic damage events.
Risk-aversion, in such a situation, would command paying a, potentially significant, positive surcharge on the SCC in order to not only reduce the mean, but also the variance of losses through emission abatement. 
Such an argument is in line with the so-called \textit{precautionary principle}, defined by \citet{sunstein_irreversible_2006} as \blockquote{\textit{when regulators are dealing with an irreversible loss, and when they
are uncertain about the timing and likelihood of that loss, they should
be willing to pay a sum – the option value – in order to maintain
flexibility for the future.}}
However, \citet{kousky_risk_2011} do not provide a preferred mechanism by which a risk premium should be calculated.

Finally,~\citet{weisbach_should_2012} counters that the precautionary principle was not applicable to Pigouvian taxes and that government shall charge simply the ``expected marginal social harm from an activity, adjusted in each period to reflect new information.''
\citet{weisbach_should_2012} justifies this conclusion by arguing that 
\blockquote{\textit{[...] environmental taxes decentralize decision-making. They impose a price on harm-causing activities and let individual actors decide the appropriate level of activity given that price. Individual actors will take precautions if appropriate. We can think of Pigouvian taxes as completing the market, and like in other situations with complete markets, we rely on individuals to determine the appropriate level and timing of their activities. The government should not additionally accelerate or delay environmental taxes.}}

This conclusion relies on a number of limitations.
First, any estimation of marginal harms would need to be based on knowledge about to what extent individual actors take precautions.
However, it is entirely unclear to what extent precautionary attitudes can be predicted in advance: even in the absence of climate change effects, precautionary attitudes can change rapidly in response to unprecedented events, as evidenced by the COVID-19 pandemic \citep{thoma_cognitive_2021}. 
Second, taxes are required to ``adjust to new information'' such that estimates of damages from earlier emissions can be revised, and thereby adjusting the future behavior.
However, given the global nature of the negative externalities associated to carbon emissions, it is not clear if this information is readily available in a setting with multiple actors making decisions related to emissions.
In other words, if assessing the impact of previous actions on total emissions and damages is difficult, then it is also nontrivial to adjust future emissions without some form of global cooperation or information sharing. 
Third, \citet{weisbach_should_2012} notes that their analysis relies on the price on pollution not to affect the pace of technological development.
Finally, the authors do not account for a large proportion of unskilled market participants like small and medium-sized enterprises (SMEs) that are subject to the \textit{Tragedy of the Horizon} and lack resources (e.g., qualified staff) to assess climate risks.

As SCC models cannot be guaranteed to be free from \textit{epistemic errors} even in the absence of \textit{environment (aleatoric) uncertainty}, \textbf{we propose} that, in the context of Pigouvian taxation, \textbf{the precautionary principle does indeed need to be applied to SCC estimates}.
Unlike well-known prior criticism of established SCC models, including DICE and RICE \citep{Stern2021}, we base our proposal primarily on the observation that the current \textit{model innovation process} for integrated assessment models (IAMs) is not optimal. 
This is immediately the case due to the lack of a reliable and transparent mechanism to assess \textit{retrospective global climate damages} that prohibits the \textit{calibration} of any SCC models to real-world data, as well as the lack of an incentive to develop such a mechanism. 
However, even if such a mechanism becomes available, we argue that the innovation process of SCC models should not rely solely on academic innovation by a small group of intellectuals, but rather on decentralized decision-making by skilled market participants - a central tenet in neoclassical economic theory \citep{weisbach_should_2012}. 

Assuming the existence of a commonly agreed methodology for measuring retrospective marginal damages, at first glance, a fair way of charging polluters would therefore be through establishing a long-term liability where polluters are periodically charged with the marginal damages incurred so far. 
According to neoclassical economics, this liability should, in general, incentivize decentralized decision makers to take future climate risks into account \citep{weisbach_should_2012}. 
In contrast, Pigouvian tax is generally interpreted as a one-time charge at the time of pollution. 
We note that there are two important practical reasons why polluters should be charged in advance.
Firstly, due to the short decision-making horizons of polluters, it is not guaranteed that they will price in their long-term liabilities adequately, thereby hampering the mitigation of long-term climate risks.
Second, compensation from polluters often needs to be pooled-in to finance long-term investments into adaptation and mitigation measures.

Thus, in the subsequent sections, we introduce our proposed mechanisms to address the aforementioned disadvantages of one-time fixed carbon taxes and difficulty of estimating the SCC.

\section{Retrospective Social Cost of Carbon Updates (ReSCCU)}
\label{sec:resccu}

\subsection{The SCC is a moving target}
The marginal cost of future damages due to an extra ton of CO2 is difficult to estimate.
This is in part due to the fundamental aleatoric uncertainty in response of the physical world to future emissions and to the greater uncertainty on future decision-making as well as socioeconomic pathways.
Since the marginal damage estimates form the basis for policymaking, there is a feedback loop tying today's estimate to tomorrow's created damage. 
For example, overestimation of today's marginal damages may inflate the expectation of future damage estimates, thereby causing harsher policies to mitigate adverse effects resulting in lower than expected future damage.
The very existence of such feedback effects may already pose questions whether a \textit{marginal damage} is indeed a well-definable concept, if not inherently tied to rigorous constraints on the nature of future decision-making processes.

However, marginal damages are hard to estimate even when considering retrospective emissions.
Assuming a polluter emits an extra ton of CO2 at time $t$ and abates an extra ton of CO2 at future time $t'>t$, how exactly do we estimate the marginal damage caused even if the pollution event is assumed confined entirely to the interval $[t,t']$?
In order to understand this question better, we need to understand how climate damages occur. 

The mechanisms through which climate damage occurs are numerous. 
Among physical mechanisms, one may further distinguish between damages from extreme weather events, such as floods, hurricanes, and droughts, and events with more permanent effects, such as regional climate changes and sea level rise. 
Estimating damages from extreme weather events may seem comparably easy, given insurance and reinsurance companies do provide annual estimates of claims incurred. 
However, global insurance coverage is patchy and, particularly when considering livelihoods, does not cover many of the world's most vulnerable communities. 
In addition, not every extreme weather event is due to climate change. 
Even estimating last year's marginal damages from emitting an extra ton of CO2, the year before, is therefore difficult and relies on the emerging areas of climate attribution science \citep{otto_attribution_2016} and global quantitative estimates, such as those increasingly derived from automated earth observation \citep{jimenez-jimenez_rapid_2020}.

Retrospective attribution of damages from sea level rise or regional climate change is equally difficult, and, apart from damages to physical assets, likewise requires inclusion of hard-to-quantify damage to human health (including psychological effects), biodiversity loss, and even quality of life. 
With data being unavailable, past attempts to quantify climate damage have largely relied on statistical methods comparing the economies of different climatic regions. 
Such approaches, however, inherently suffer from numerous confounding effects. 
A central problem concerns the costs incurred during adaptation.

With climate damage hard to quantify retrospectively, extrapolating it into the future is naturally even harder. 
Firstly, the future transient climate response to rising levels of GHGs, while comparatively well-understood, inherently carries uncertainty. 
Secondly, future damages incurred depend on the particularities of future economies. 
We already noted how a future economy's adaptability and resilience to climate change influences the levels of damage incurred with respect to its physical effects, such as increases in the frequency of extreme weather events. 
Both these trends are compounded and inherently interlinked through the presence of countless decision makers in the system dynamics whose behavior, even if their incentives and policies were fully known in advance, might readily drive system dynamics virtually unpredictable due to the emergence of bounded rationality effects \citep{botzen_bounded_2009}, system bifurcations such as those emerging from social tipping points \citep{otto_social_2020}, and, fundamentally, chaos arising naturally in multiplayer games \citep{sanders_prevalence_2018}. 
These fundamental constraints are largely ignored in current IAMs (see Section \ref{sec:iam}): For example, in DICE \citep{nordhaus_revisiting_2017}, population growth is treated as an exogenous variable - rather than an endogenous variable that should depend both on social dynamics as well as environmental constraints and effects \citep{stern_time_2021}.

We have thus established that marginal damages from GHG emissions are intrinsically hard, or even impossible, to predict across sufficiently long time horizons given the numerous sources of \textit{aleatory}, i.e., environmental uncertainty. 
Even under optimistic assumptions, such as predictable decision-making, exploiting any predictability poses hard \textit{epistemic}, i.e., modeling challenges: First, a lot of the required empirical data, even if theoretically measurable, is not currently available, and can often only be made available at significant cost. 
This issue ranges from a lack of weather and climate data in areas such as West Africa, up to insufficiently fine-grained economic data. 
Secondly, even if the necessary data was available, models would need to be run at sufficiently high resolution and scale, thus incurring complex model design and calibration challenges, as well as substantial computing costs. This emphasizes that significant financial incentives are required for model innovation.

\subsection{Formal Definition}

In the wake of the above discussion, we introduce Restrospective SCC Updates (ReSCCU), to formalise the idea that SCC evaluations can, especially close to the time of emission, only ever be regarded as preliminary estimates.
We use the notation $\widehat{\text{SCC}}_{t}^{t_0}$ to denote the estimated SCC of one tonne of emission occured at time $t_0$ but evaluated at time $t > t_0$.
ReSCCU posits that, after emitting a unit quantity of GHG emissions at time $t_0$, any initial estimate of the SCC, $\widehat{\text{SCC}}_{t_0}^{t_0}$, should be seen in anticipation of periodic future ReSCCUs, $\Delta_t^{t_0}$ at time $t$.
Thus, the \textit{true SCC} for emissions at time $t_0$, denoted by $\text{SCC}_{t_0}$, accounting for the present value of the sum total of future climate damages, is given by

\begin{equation}
\label{eq:scc}
    \text{SCC}_{t_0} = \widehat{\text{SCC}}_{t_0}^{t_0} + \sum\limits_{t=t_0+1}^{t=\infty}\underbrace{\widehat{\text{SCC}}_t^{t_0} - \widehat{\text{SCC}}_{t-1}^{t_0}}_{\Delta_t^{t_0}}
\end{equation}

Note that as the time since emission advances, our estimate of the true SCC becomes increasingly dominated by \textit{empirically measured} retrospective, rather than forecast, climate damages.
In other words, as $t - t_0$ becomes larger, $\widehat{\text{SCC}}_t^{t_0}$ is increasingly dominated by retrospective damage estimation. 

By requiring SCC estimates to always be considered jointly with future retroactive adjustments, ReSCCU, at first glance, may seem to undermine the very utility of having SCC estimates, i.e., having access to a single number that can summarize a unit emission's future climate damages here and now. 
A more subtle consideration, however, reveals that ReSCCU reduces to the traditional notion of SCC precisely when the discounted future damage estimation functions are exact, and deviates from the SCC estimates to the degree in which they are not. 
If we assume that our SCC estimation functions are largely correct, then we expect the ReSCCU adjustments not to be significant.

While ReSCCU acknowledges the inherent uncertainty of the SCC extending indefinitely into the future, it remains to be seen how this hypothetical and idealized concept can be implemented in practical settings.
As a first step in that direction, we need an independent and transparent modeling agency that can release the annual SCC estimates. 
We call such an institution the \textbf{RetroAgency}.
To ensure fair regulation, we propose a governing structure of the RetroAgency to follow the guidelines of similar institutions such as IPCC.
In addition, SCC estimation mechanisms may evolve with time, but this will only affect future models.
As a result, the RetroAgency should be transparent about the models and the assumptions used each year. 
A RetroAgency is tasked with the major responsibility of releasing their SCC estimates for previous years. 
For example, at time $t$, the RetroAgency will be tasked to release their past $n$ years of estimates, i.e., $\widehat{\text{SCC}}_t^{t}, \widehat{\text{SCC}}_t^{t-1}, \widehat{\text{SCC}}_t^{t-2}, ..., \widehat{\text{SCC}}_t^{t-n}$. 
These estimates can then be used to compute annual ReSCCUs. 

In the following sections, we propose two market-based mechanisms to implement PPP-conformant schemes through leveraging ReSCCU as established by RetroAgency.

\section{Retroactive Carbon Pricing (ReCaP)}\label{sec:recap}

In this section, we discuss \textit{Retroactive Carbon Pricing (ReCaP)}, a concrete mechanism that seeks to implement ReSCCU in the context of carbon taxation with the help of participation from the government.
Our novel mechanism design incentivizes the continuous innovation of retrospective climate damage estimates. 

\begin{figure}
    \centering
    \includegraphics[width=1.0\linewidth]{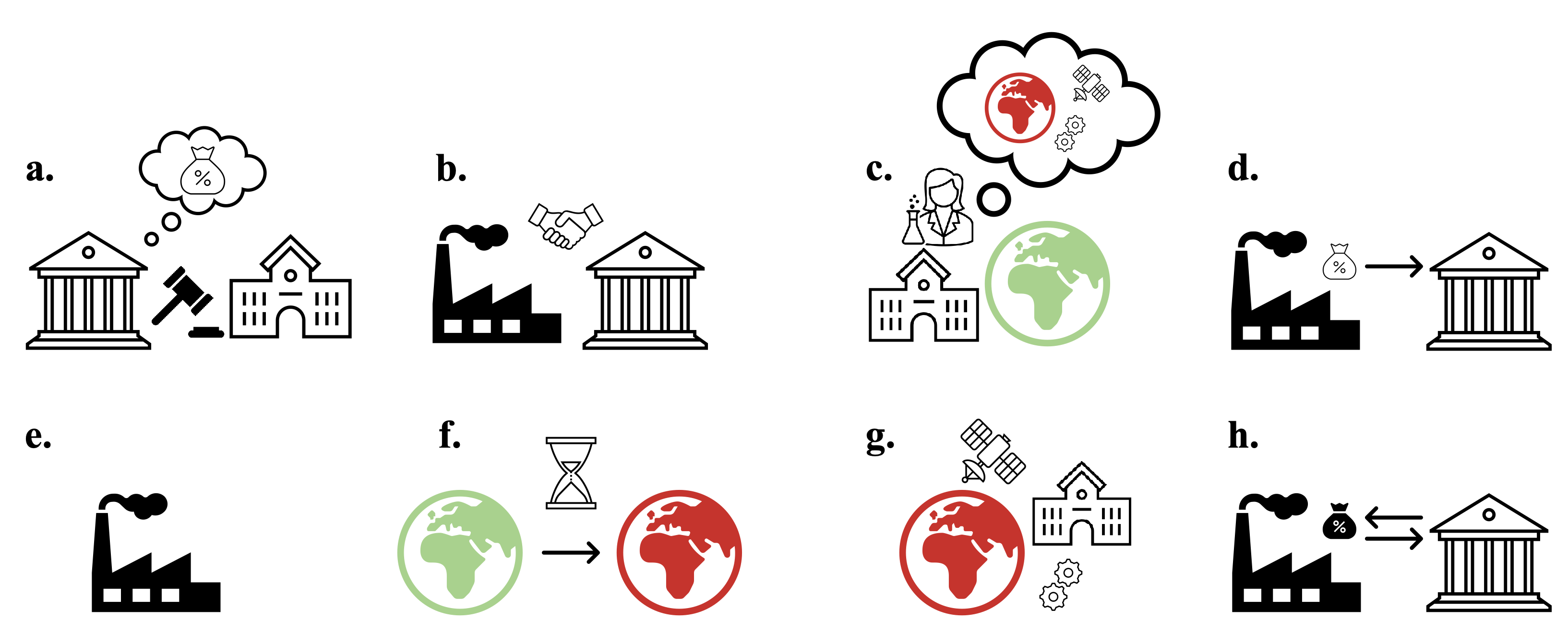}
    \caption{Schematic depiction of idealized ReCaP. a) The government enacts a retroactive carbon tax law and establish an independent and transparent RetroAgency b) Polluters and government agree on next year's emissions volume c) RetroAgency establish the best present SCC estimate d) Government taxes polluters according to the SCC estimate e) Polluters pollute while f) a year passes. g) Real-world climate damages from the past year are measured and adjustments to past year's SCC estimate are calculated by the RetroAgency h) Government and polluters settle SCC adjustment payments (also periodically over the coming years).}
    \label{fig:recap_basic}
\end{figure}

Ideally, ReCaP works as follows (see Figure \ref{fig:recap_basic}): First, the government enacts a retroactive carbon tax law that details the ReSCCU mechanism to adjust the carbon tax over time.
The government also establishes an independent and transparent RetroAgency.
After the carbon tax law comes into effect, polluters need to register their expected CO2 emissions for the ensuing year and pay the associated ReSCCU estimate according to the mechanism enshrined in law. 
Going forward indefinitely, after each additional year has passed, the government and the polluters exchange cash flows according to the ReSCCU adjustments.
This straightforward incorporation of ReSCCU into Pigouvian taxation does, in principle, expose polluters optimally to the best possible SCC estimation in the presence of epistemic uncertainty. 

\begin{figure}[ht]
    \centering
    \includegraphics[width=1.0\linewidth]{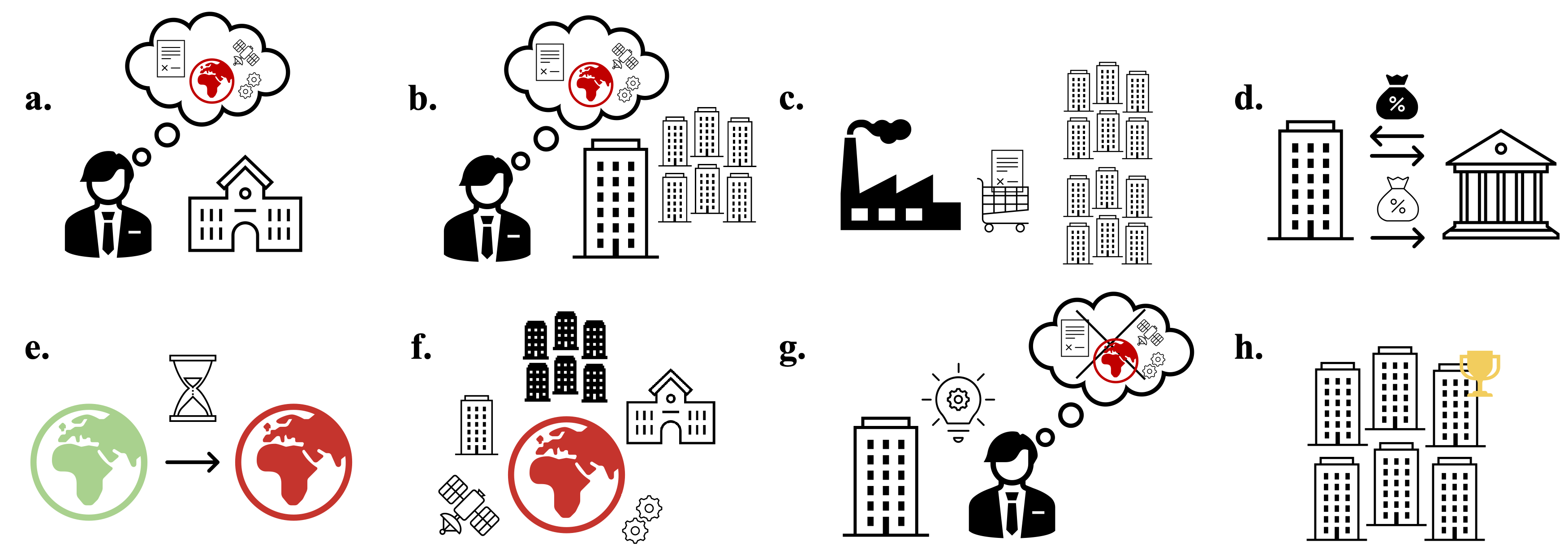}
    \caption{ReCaP with swap contracts between the insurers and the government. (a) An independent RetroAgency such as IPCC estimates the SCC. (b) Financial institutes create SCC models to estimate the right premium for a policy product. They will compete on the swap market. (c) The polluter chooses the best available insurance policy product to insure its future pollution. (d) The swap contract now takes over any ReSCCU adjustments between the financial institutes  and government. (e) A year passes. (f) Retrospective climate damages are measured by RetroAgency, each financial institute may do this differently. (g) The financial institute innovates its models according to the new data. (h) The financial institute with the best SCC model can offer the lowest premiums for the swap.
}
    \label{fig:recap 2}
\end{figure}

However, the above scheme has the following drawbacks: (a)  If the polluters go bankrupt sooner than the time period considered for ReSCCU, the polluters will not be paying the full cost of damages.
This can incentivize the participants to game the system. 
(b) How can individual polluters, in particular SMEs, manage the risks associated with variable ReSCCU adjustments?
Such companies do not have the skill, the expertise, the data and the economies of scale to do predictive modeling of future climate damages caused by their current emissions.
Thus, to safeguard against these shortcomings, we introduce the insurance companies to take over the risks of annual ReSCCUs (see Figure \ref{fig:recap 2})

An insurance company uses its existing skills and economies of scale in order to establish an SCC prediction model to estimate the right premium ahead of time, which should be equal to $\left(\text{SCC}_{t_0}-\widehat{\text{SCC}}_{t_0}^{t_0}\right)$ plus the model innovation costs and the desired profit. 
Importantly, ReCaP requires that several competing insurance companies undergo this process independently, creating a market for ReSCCU insurance policies available to polluters.
Polluters are now able to buy insurance against ReSCCU adjustments for their respective emissions on this market.
This makes polluters liable for a one-time insurance premium at the time of pollution, after which they are no longer exposed to the risk of ReSCCU adjustments. 
Instead, \textit{the insurance companies settle any future cash flows with the government directly through a swap contract, overcoming the Tragedy of the Horizons by offloading risks to skilled market participants whose business models have always relied on long-term horizons.}

\subsection{Opportunities}

Besides addressing the issues stemming from the Tragedy of the Horizon, ReCaP offers the following additional benefits.

\paragraph{Model innovation.} 
The need for ReSCCU is motivated by the fact that the SCC is hard to estimate at the time of emission. 
While some of this difficulty arises from irreducible aleatoric environmental uncertainty, current SCC models almost certainly suffer from epistemic uncertainty, i.e.,  current models do not optimally process all relevant data that already is or could be available. 
The reason for this is clearly that there are insufficient incentives for model developers to create alternative IAMs, or improve existing ones through either model innovation and/or additional data collection. 
While it is intrinsically unclear where exactly the boundary between aleatoric and epistemic modeling error lies, ReCaP can be seen to provide substantial financial incentives to insurance companies to continuously improve their ReSCCU forecasting models.

As insurers compete in the market, the insurance premia will eventually converge to the marginal costs, which are likely dominated by ReSCCU $\Delta_t^{t_0}$ adjustments. 
This means that insurance companies are incentivized to continuously innovate their models such as to reduce their forecasting error relative to future ReSCCU adjustments, which benefit from hindsight. 
Thus, insurance company surcharges arising in ReCaP can be interpreted as a precautionary investment in technological innovation, which, as other technological investments (e.g., renewable energy production), are justified even from a decentralized market perspective \citep{weisbach_should_2012}.

\paragraph{Polluter-Pays Principle perspective.} Uncertainties surrounding greenhouse gas pollution also pose a number of questions when considering the implications of PPP\citep{tokuc_rio_2013}, which is a statute of international environmental law that posits that 
\blockquote{\textit{National authorities should endeavour to promote the internalization of environmental costs and the
use of economic instruments, taking into account the approach that the polluter should, in principle,
bear the cost of pollution, with due regard to the public interest and without distorting international
trade and investment.}}
Isn't it unfair if polluters are charged according to estimated damages, rather than actual damages incurred? 
And should polluters be held responsible for suboptimal future decision-making? 
For example, if a future government incentivizes house building near the shore, naturally resultant property damages due to sea level rise may increase. 
As future decision-making could be arbitrarily ignorant, can we really make a polluter responsible for it at the time of emission? 
If not, what kind of future decision-making is the polluter liable to? By basing Pigouvian taxation rates on the ReSCCU rather than the SCC, many of these questions can be addressed optimally.

\paragraph{Niche businesses and increased funding for climate research}

With model innovation playing a central role in the ReCaP market, we may observe appearance of new business models, increase in the activity of relevant businesses, or more funding for academic researchers either through industry-academia partnerships or directly through government funding. 
As climate science plays a central role in building these models, it is imperative that climate modeling could itself become a bigger business model.
Due to the dependency of climate modeling on various modalities of data, emergence of data collection government initiatives or businesses might become necessary. 
In effect, there will have to be sufficient funnelling of finances to enhance climate modeling so that it is viable for insurers to thrive.

We note that the very issue that ReSCCU is set to address, i.e., bringing the future to the present, might also be a motivating factor for some climate modeling businesses to act unethically via creating superficial models for the sake of sales. 
This is made feasible as their predictions cannot possibly be verified until a decade has passed, during which these businesses might want to exit the market.
In order to prevent such activities, we propose RetroAgency to provide minimum guidelines for any such climate modeling businesses. 
To ensure such guidelines have been implemented, we might see an emergence of climate modeling auditing businesses to ensure that the models pass the minimum prescribed standards. 

\subsection{Challenges}

Finally, we note that the ReCaP scheme also poses a few practical challenges related to risk management, as well as political buy-in and initial hurdles to implementation.

\paragraph{The limits of risk mitigation.} We have so far assumed that insurance companies can indeed estimate the total future ReSCCU adjustment, $\Delta_t^{t_0}$, well enough in order to effectively price in risks through one-time premia at the time of CO2 emission. 
In fact, there are two main challenges in doing so: First, it is inherently unknown to what extent $\Delta_t^{t_0}$ can be predicted and whether it is indeed bounded. 
Second, insurance companies may find it difficult to diversify their climate risks, as climate disasters tend to be correlated across regions and asset classes.
Various measures could be taken in order to offset risks to insurance companies.
Perhaps most straightforwardly, reinsurers, or, ultimately, governments, could choose to bear excess tail risks. 
This could be implemented by choosing a finite time horizon for $\Delta_t^{t_0}$, introducing a  temporal discounting rate or simply through side contracts. 
While such measures would be successful in managing risk to insurers, they would at the same time dilute the efficacy of the Pigouvian taxation mechanism. 

If the total insurance volume for ReCaP transactions becomes sufficiently large, the economic survival of insurance companies may start to be regarded as vital to financial and economic stability (``too big to fail"). 
This might dilute the perceived risks that insurance companies feel exposed to as they may bet on ultimately being bailed out by the government, potentially leading to artificially low premiums.

\paragraph{Political buy-in and hurdles to practical implementation.} As a carbon taxation mechanism, ReCaP requires political buy-in from the governments and insurance companies. 
Currently, operating carbon taxation schemes tend to underprice CO2 emissions from a climate damage perspective, perhaps indicating limited political support for ReSCCU-like adjustments.

\section{Private ReCap (PReCaP)}

\label{sec:precap}

In Section~\ref{sec:recap}, we introduced ReCaP as an application of ReSCCU to Pigouvian taxation. 
While, in theory, ReCaP improves over SCC-based taxation, it also poses a few practical challenges grounded in risk diversification, as well as large-scale political buy-in and systemic relevance. 
In face of these obstacles, and as a complementary measure, it does serve well to consider: What can private investors do in order to help implement ReSCCU?

In this section, we introduce \textit{Private ReCaP (PReCaP)}, a novel mechanism that enables ReSCCU to be be implemented with minimal government participation.
Importantly, unlike ReCaP, PReCaP could, in principle, see real-world implementation based on the engagement of a few high net-worth individuals or independent institutions as well as a minimal participation from the government. 

\subsection{The PReCAP prediction market}

\begin{figure}[ht]
    \centering
    \includegraphics[width=1.0\linewidth]{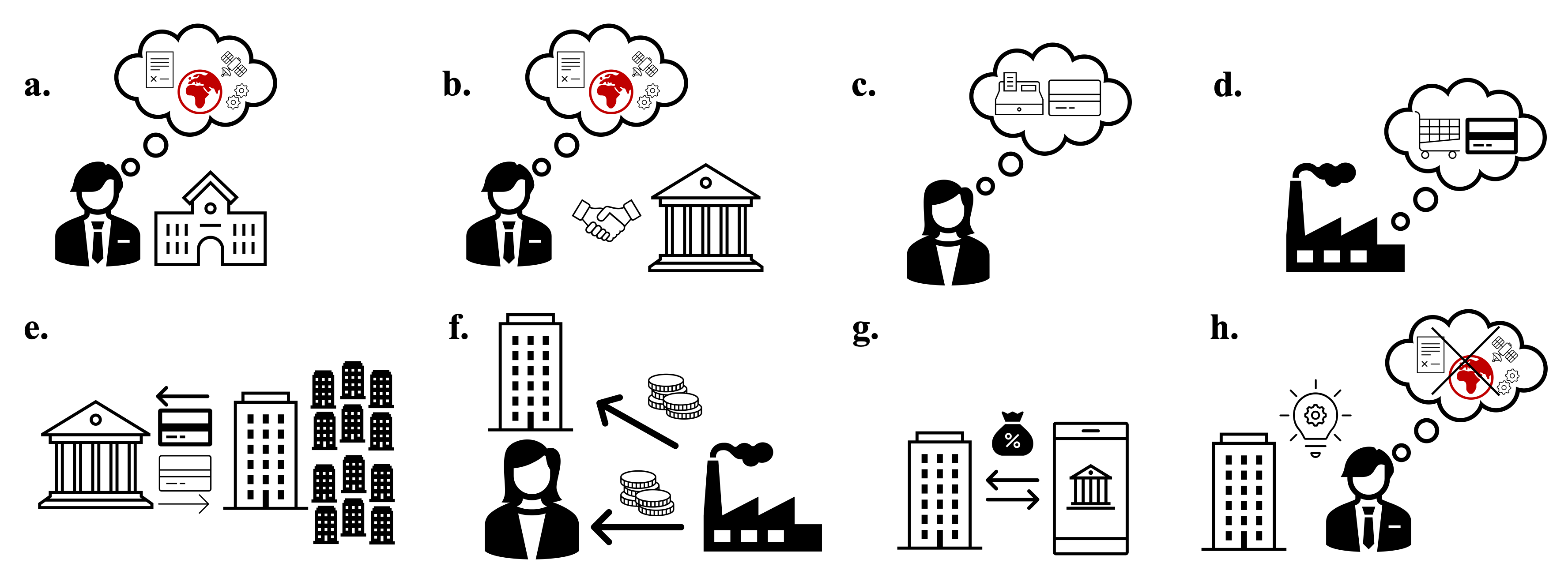}
    \caption{Private ReCaP. (a) An independent RetroAgency devises an SCC machine. (b) A RetroExchange chooses a RetroAgency’s SCC machine. (c) A seller wants to sell an uninsured carbon credit (UICC). (d) A polluter wants to buy an insured carbon credit (ICC). (e) The RetroExchange picks the best counterparty to convert the UICC to ICC. This allows buyer and seller to execute the transaction. (f) During the transaction, the polluter pays for the carbon credit as well as the insurance policy. (g) RetroExchange and the counterparty exchange PReCaP adjustments over a future time period. (h) Counter-parties undergo SCC model innovation.
}
\label{fig:precap}
\end{figure}

In its basic form, PReCaP implements competitive incentives between a group of insurance companies within the context of a \textit{prediction market} (also called \textit{betting market}) that are trying to predict future ReSCCU adjustments estimated by a trusted RetroAgency. 
This is achieved by creating a market for insurance policies in which demand is stimulated artificially by a sponsoring agency, which we dub the RetroExchange, which can, in turn, derive a decentralized market estimate of $SCC_t-\widehat{SCC}_t$ based on the spread of insurance premiums offered by the insurance companies (see Figure \ref{fig:precap}). 

\paragraph{Risk management.} The RetroExchange bears the insurance default risk. 
When an insurer goes bankrupt, this means that any associated outstanding retroactive payments will not be covered, but also that any outstanding claims will not be enacted.
Hence, insurance default may not necessarily be unprofitable to the RetroExchange.
The RetroExchange takes furthermore the risks associated with being on the other side of the swap contracts. 
This means that large negative retroactive adjustments might cause the RetroExchange to go bankrupt, in which case the insurances would bear the RetroExchange default risk. 
Such systematic risks can be lowered by traditional ways, e.g.,  requiring insurers to have reserves posted with the exchange in a default fund, or through reinsurance.  
Other ways to safeguard against such defaults can include flooring RetroAgency's cumulative pricing adjustments, which may interfere with the quality of the SCC signal, but may be a sensible way to deal with extreme value risks on the tail.
Further, RetroExchange might benefit from the revenue stream by selling the SCC signal derived from the observed insurance premium spreads, as these will only be revealed to the RetroExchange (see Appendix C for derivation of the SCC signal). 

\subsection{PReCaP as a breakthrough technology investment}

While PReCaP does not implement the polluter-pays principle for greenhouse gas emissions in full, we here discuss a way in which polluters could be charged with the costs of model innovation for SCC estimation, thus internalizing part of the mitigation costs. 
This may be achieved by implementing PReCaP in the context of voluntary carbon markets, as spearheaded by the Taskforce on Scaling Voluntary Carbon Markets (TSVCM). 

TSVCM has long recognized the centrality of the challenge to divert sufficient cash flows to breakthrough technologies, i.e. early-stage technologies that currently create carbon credits at uncompetitive prices (we dub these \textit{breakthrough credits}). 
In fact, \citet[Final Report]{tsvcm_taskforce_2021} observes that:

\begin{quote}
Many of the investments needed to scale
emerging breakthrough technologies do
not meet the risk and return expectations of
today’s markets. A range of mechanisms will
be needed to ensure capital flows to these
technologies. These could include blended
financing, access to benefit markets (including
voluntary carbon markets), or altering risk,
return or time horizon expectations for
projects with the highest potential for climate
impact. [...] For finance to flow to these GHG emissions
avoidance/reduction and removal/
sequestration projects, well-functioning
voluntary carbon markets will be a critical
enabler.
\end{quote}

As an alternative approach to these challenges, we propose to introduce regulation that would require polluters to acquire a fixed percentage of their carbon offset credits from breakthrough credit suppliers. 
To prevent market manipulation, the price of such breakthrough credits would be capped at the social cost of carbon (SCC).

In theory, such a mechanism would enable a sizable amount of cash to be diverted from polluters to breakthrough technology innovators: Even a modest $10\%$ requirement and a modest SCC of USD~$100$ per ton could, according to TSVCM predictions, result in tens of billions (and up to hundreds of billions) of USD being diverted to breakthrough technologies by 2030\footnote{Assuming a demand of 1-2 Gt CO$2$ by 2030 \cite{tsvcm_taskforce_2021}.}. 
Politically, such regulation may find substantial backing due to its ability to both efficiently tackle the breakthrough technology innovation problem, while implementing a pure form of the polluter pays principle.

The challenge with this approach is, of course, that it requires fixing a suitable SCC level as SCC estimates defined through political processes tend to be both too low from a climate safety, as well as from a volume of total breakthrough subsidies aspect. One option would be to fix the SCC in line with what the IPCC deems safe \cite{ipcc_global_2018}. 

However, this setup itself does not address the need for SCC model innovation, which we argue to be just another form of a breakthrough technology. The simplest way to incorporate SCC model innovation into our setup would be to levy a small surcharge on top of the current SCC estimate, which would then be diverted to e.g. academic grants to SCC research. However, in this paper, we have outlined that such a reinvestment may not be particularly efficient.

In the next section, we describe how, instead, ReSCCU could replace the SCC in such a setting, resulting in SCC estimates that would correct themselves over time, and subsidizing SCC model development through the creation of an SCC prediction market.

\subsection{Insuring carbon credits: A mechanism to charge for model innovation}

Having discussed the financing of breakthrough technology innovation as a core challenge to TSVCM, we now propose a detailed mechanism for how ReSCCU surcharges could be implemented in the context of a voluntary carbon market based on a form of a scalable \textit{symbolic insurance} for carbon credits. 

Central to this ambition is the realization that stock exchanges trading carbon credits, due to their pre-existing infrastructure and in-house skills, may rather naturally assume the role of a suitable sponsoring agency for PReCaP.
Additionally, polluters acquiring carbon credits from suppliers can easily be charged with ReSCCU model innovation costs if such a \textit{RetroExchange} requires each carbon credit to be \textit{insured}.
Together, this gives rise to a decentralized market mechanism that both stimulates model innovation and reveals an improved SCC estimate while being financially sustained by polluters - a proposal very much in the spirit of the polluter-pays principle.

We consider two types of sellers at the RetroExchange: (a) the carbon credit supplier, and (b) the insurers who post the premium on these carbon credits.
First, there are sellers who have acquired the rights to sell carbon credits either by the way of a project that will, in the future, or has already captured a ton of carbon.
We call these uninsured carbon credits (UICCs).
The RetroExchange asks the insurance companies for a premium at which they could insure UICCs, following which the exchange selects the offer that provides the lowest premium. 
\textit{The optimal estimation of the SCC is the lowest price that still allows the insurer to make money. 
Hence, competitive bids combined with the need of the insurer to still make a profit will help favor those insurers who develop more accurate models.}
UICCs along with the insurer chosen by the RetroExchange, form an insured carbon credits (ICCs).
Buyers bid on the RetroExchange for ICCs; UICCs are not directly purchasable. 
Finally, the selected insurer enters into a swap contract with the RetroExchange, thereby exchanging the ReSCCU adjustments on a periodic basis for a fixed period of time.

Insurers operate on the sell side of the exchange as they post the asked insurance premium and volumes of insurance contracts they are willing to sell. 
For simplicity, let us assume that all retroactive payments are capped to $30$ years since conversion from UICC to ICC\footnote{Otherwise we need to distinguish between ICCs of different maturities}.
The clearing mechanism by the RetroExchange is now an auction with a complication: 
While ICCs can be cleared straightforwardly, UICC clearance involves a form of transaction cost (or friction) due to the need to match bids and asks simultaneously with the lowest available policy insurance option. This transaction cost could be automatically applied by the exchange, which would add the insurance costs to the spread. In principle, this form of ``three-agent" transaction could be decomposed into two 2-agent transactions, where the buyer needs to present a previously acquired insurance certificate at the time of UICC acquisition. This, however, would allow polluters to strategically hedge insurance certificates, which may introduce a distorting effect on insurance policy pricing. A more detailed discussion of such an auction mechanism is beyond the scope of this paper.

\subsection{Other incentives for participation}

A crucial aspect of PReCaP is that the participation in it is \textit{voluntary}, meaning that additional incentives are required in order for carbon credit buyers and sellers to be willing to pay for insurance surcharges. 
We argue that the following factors may plausibly incentivize voluntary participation in PReCaP:

\begin{itemize}
	\item \textbf{Corporate social responsibility (CSR):} It is conceivable that a number of large corporations would be willing to acquire a small share of their carbon credits through the RetroExchange for reasons of corporate social responsibility. This may incentivize polluters to expose at least a small share of their carbon credits to the PReCaP scheme.
	\item \textbf{Supplier support:} It is conceivable that a number of carbon credit suppliers may choose to exclusively sell through the RetroExchange for CSR (corporate social responsibility) or marketing reasons. As demand for carbon credits outstrips supply, this would then incentivize polluters to acquire any credits available, even if subject to an insurance premium. 
    \item \textbf{Carbon credit pricing:} Institutional lenders, such as banks, require access to accurate long-term interest rate benchmarks. Such estimates would likely benefit from improved SCC forecasts. This may encourage banks to invest in PReCaP implementations.
	\item  \textbf{Risk Diversification}: In the literature on catastrophe risk it is shown that catastrophe risk insurance as an investment has small correlation to other market indices \footnote{Though it is unclear if this would hold for climate risks due to pollution as well, since there is a causal link between economic activity and carbon outputs.}. This may form financial incentives for insurance companies to participate in PReCaP schemes.
	\item \textbf{Regulatory incentives:} Governments, or supragovernmental organisations, could, in principle, introduce regulation incentivizing participation - for example by requiring all newly released voluntary carbon credits to be ReSCCU insured at first transaction -, or directly investing in, PReCaP implementations - this would require much less political buy-in than ReCaP, which, in addition would be much more easily scalable.
	\item \textbf{Charity and Philanthropy:} High net-worth individuals alone could subsidize small-scale PReCaP schemes. 
\end{itemize}

\section{Discussion}
Our proposal and discussion in this paper aim to draw the attention of policymakers, climate scientists, computer scientists, and economists to an alternative solution to inefficient resource allocation and risk assessment to address climate change.
We invite public, private and academic communities to discuss the weaknesses and strengths of our approach, with the goal of morphing it into even more powerful mechanisms that can potentially help address the challenges related to the greenhouse gases market failure, climate financing, and climate risk assessment.
We highlight below a number of directions for future research that will require considerable concerted international efforts.

\paragraph{Toward ReSCCU: Implementing (P)ReCaP}
We propose to replace the SCC by ReSCCU in all relevant contexts, ranging from policy evaluation to carbon taxation through ReCaP.
To this end, we argue for concerted efforts by a few aligned high net-worth investors in a small-scale, limited agent-based real-world implementation of PReCaP, as well as an exploration into integrating PReCaP in the context of voluntary carbon markets. 
Such a \"sandbox\" market would stimulate model innovation and reveal a decentralized market estimate of the SCC that could be used as a universal benchmark signal.

\paragraph{Learning from physical climate modeling: Toward an IPCC for SCC modeling}

Like IAMs, physical climate models, i.e., models that establish the physical response of the earth's climate system to greenhouse gas emissions, likewise need to incorporate numerous aleatoric and epistemic uncertainty sources. 
The scientific community has responded to this by considering ensemble predictions based on curated ensembles of independently developed physical climate models.
In this way, international scientific organizations, such as the IPCC can help reduce epistemic uncertainty through exploiting model diversity. 

Unfortunately, the diversity among leading IAMs is currently rather small, perhaps both owing to a lack of funding in academic IAM research, as well as the highly politicized nature of IAM adoption by governments.
We therefore suggest that funding in IAM research be increased in order to tackle epistemic modeling errors.
At the same time, the enormous challenges in data collection for IAM calibration and validation on a global scale, as well as the need to generate trustworthy conclusions calls for the establishment of an international institution for SCC estimation.

\paragraph{AI as a superpower: Fostering data collection, model innovation, and diversity}
We suggest that next-generation IAMs can benefit from recent breakthroughs in artificial intelligence, large-scale data collection and exa-scale compute to a similar extent that physical climate models already do. 
In particular, rapid advances in large-scale agent-based modeling, machine-learning driven calibration and validation methods, including emulation, inference and multi-agent learning approaches, offer attractive avenues toward achieving significant model innovation in the IAM space.

At the same time, we remark that the lack of a transparent methodology to derive even global estimates of \textit{historic} climate damages is a serious obstacle towards the construction of any trustworthy IAMs for SCC estimation, yielding a lack of data for empirical backtesting. 
This concern affects any currently known IAM, including DICE, RICE, and PAGE.
At the same time, this insufficiency exemplifies the need to introduce model innovation incentives to the IAM community, as many recent advances in other scientific areas, including methodology advances in AI, as well as, e.g., advances in extreme weather attribution in the physical sciences, are clearly ignored in current generations of IAMs. 

\paragraph{SCC model innovation as climate technology innovation.} Even if we achieve global net-zero emissions by 2050, subsequent decades will require ongoing negative emissions in order to stabilize warming at or below 1.5$^{\circ}$C \cite{ipcc_global_2018}. 
Many of the future technologies required to achieve this currently have \textit{breakthrough} status, meaning they generally have not been proven at scale and aren't cost competitive with other existing approaches, such as reforestation or peatland restoration. 
Many academics and international organisations have acknowledged that diverting sufficient cash flows to subsidize breakthrough technology innovation is a central challenge to climate mitigation \cite{ipcc_global_2018,tsvcm_taskforce_2021}. 
In this paper, we propose that SCC model innovation and breakthrough technology innovation are indeed intrinsically connected: by ensuring that a certain proportion of breakthrough offsetting credits are acquired by polluters at the true SCC rates every year, we would provide a possible solution to the breakthrough technology problem. Investments in SCC model innovation, e.g. by implementing ReSCCU through the PReCaP scheme, are in fact critically required rather than a wishful addition.

\paragraph{Escaping from ideological trenches}
Lastly, we do not advocate for or against using social cost of carbon-approaches over target-consistent ones. 
The latter, while suffering from a variety of weaknesses related to climate justice in the face of remaining carbon budgets, may in fact be a more sensible approach in the face of extreme tail risks, including an end to human civilization as we know it.
However, we do suggest that if the social cost of carbon is to be used, either solely or in combination with other indicators, then it is imperative that it be thought of through the lens of ReSCCU. 
Both target-consistent and SCC-based approaches have different advantages and disadvantages, suggesting both might be used together rather than in isolation, and therefore justifying investment in further researching both.

\paragraph{ReSCCU for Cap and Trade schemes} We note that ReSCCU could potentially be applied in order to retroactively adjust the price floor of cap and trade carbon markets. The resulting legal and implementational challenges are likely similar to the ones discussed for PReCaP. Adjusting the EU ETS price floor has been extensively discussed and modeled as a possible policy intervention for adjusting price distortions due to interactions with exogenous green policy subsidies \cite{richstein}.  One way to reconcile ReSCCU with cap and trade markets could thus be to retroactively adjust the carbon market price floor. We leave further examination of such a ReSCCU implementation to future work.

\section{Conclusion}
\label{sec:conclusion}
In this paper, we introduce a new scheme to internalize negative externalities due to climate change. In contrast to the Pigouvian tax, our Retroactive Social Cost of Carbon Updating (ReSCCU) scheme levy the taxes retrospectively so that every dollar accounts for the long-term impact from CO2 emission to the environment.

As a basic implementation of ReSCCU, we proposed the Retroactive Carbon Pricing (ReCaP) mechanism which requires polluters to only pay the sum of SCC estimate for that year (put forth by an independent and transparent agency) and a premium offered by a selected insurer. 
In doing so, not only do we facilitate polluter pays principle making polluters the correct social cost of carbon, but also address the default risk of the polluters and enable the participation from polluters who lack the ability to do sophisticated modeling to estimate the SCC.
Large financial institutions enter into swap contracts with the government exchangin ReSCCU peridoic adjustments.
With advantages on information flows and benefit from economies of scale, are incentivized to innovate their SCC estimation models.

As a successful ReCaP requires political buy-in, which can be a hurdle given the short operating horizon of governments, we further proposed to seed the machinery for ReCaP with a Private ReCaP (PReCaP) mechanism which implements ReSCCU in the context of a private voluntary carbon market. 
PReCaP retains advantages of ReCaP, such as fostering model innovation and implementing the polluter pay principle, while also allowing to derive an improved, decentralized market estimate of the true SCC.

Our effort displays a novel approach in the climate change literature using market-based mechanisms to address a deeply concerning bottleneck in the creation of integrated assessment models by fostering model innovation, by creating a prediction market for insurers (ReCaP) or suitably skilled private counterparties (PReCaP). 
By designing such a market carefully and excluding unskilled participants, we propose to overcome the Tragedy of the Horizon inherent to previous decentralized approaches toward SCC estimation.

Viewing ReSCCU as a special form of climate technology innovation creates a particularly powerful synergy: by charging a fraction of breakthrough carbon credits at ReSCCU prices, voluntary carbon markets could optimally incentivize both breakthrough climate technology innovation and SCC model innovation.

\section{Contributions}
Y.B conceived the original idea of ReSCCU and ReCaP as well as supervised all aspects of the research.
C.S.d.W drew connections of the original idea of ReCaP to Pigouvian taxation, as well as proposed and developed PReCaP.
C.S.d.W also noted the need for a global effort to establish proper estimates of retrospective carbon damages, and provided the project’s overall climate policy framing.
C.S.d.W, M.S, A.W, T.Z, P.G, and Y.B discussed the market mechanisms in PReCaP.
C.S.d.W, Y.Z, A.W, T.Z, M.S, and D.R did a full literature review to draw the connections of ReSCCU to the existing literature.
C.S.d.W, D.R and T.Z made the figures and the plots presented in this paper.
A.W and T.Z did a thorough literature review on the drawbacks of IAMs to estimate SCC, and, with the help of C.S.d.W, M.S, Y.Z and P.G, expanded our discussions on limitations of cap-and-trade as well as market-based mechanisms to estimate SCC.
T.Z and M.S helped C.S.d.W further the understanding of drawbacks of Pigouvian Taxation.
M.S and T.Z provided analysis on the swap formulation of ReSCCU updates.
M.S raised the issues of insufficient SCC model diversity that centrally inspired the PReCaP approach.
Y.Z brought the macroeconomic and policy design perspective to the group discussions, as well as motivated the integration of PReCaP into voluntary carbon markets.
The initial manuscript was written by C.S.d.W, Y.Z, A.W, M.S, D.R, T.Z, and P.G.
P.G setup the team, facilitated the discussions, structured the white paper, and, with the help of C.S.d.W, managed the direction of the project.
All authors have devoted time and energy to brainstorm all the ideas in the paper.
Final manuscript has been read and revised by all the authors.

\section{Acknowledgements}
Authors want to thank Mark Carney for his feedback and suggestions on aligning PReCaP for breakthrough technology to stimulate technology innovation (section 7.2).
Authors are also thankful to Don Coletti, Cameron Hepburn, Vincent Conitzer, and Francois Mercier for their inputs on various aspects of the project that have helped us in shaping the ideas presented in this paper. 
Y.B, P.G and T.Z are grateful for funding from CIFAR, NSERC and Microsoft.

\bibliographystyle{plainnat}
\bibliography{references}

\appendix

\section{Existing Carbon Pricing Approaches}

\subsection{Estimating the Social Cost of Carbon With Integrated Assessment Models} \label{sec:iam}

Policymakers and institutions heavily rely on integrated assessment models (IAMs)—models that simulate the interactions of human decision-making about the economy, energy systems, and land use with the natural world—to produce estimates of the SCC \cite{UNFCCCIntegratedModels}. 

IAMs generally have relatively simple representations of the economy and don’t have a financial system. 
They also don't give fiscal and monetary policy responses that might help smooth economic fluctuations. 
This type of model examines the economics of climate change through the lens of neoclassical economic growth theory, in which emissions are viewed as negative natural capital and emissions reductions as investments.
According to these models, devoting output to abatement activities reduces consumption today while preventing economically damaging events related to climate change. 
As a result, future consumption possibilities are expanded. 
The overall assessment is highly dependent on the value of the discount rate.

DICE (Dynamic Integrated model of Climate and Economy) and RICE (Regional Integrated model of Climate an Economy) models are some of the most known IAMs. 
They were developed primarily by climate economists, led by the Nobel Prize winner William Nordhaus.
DICE integrates a general equilibrium model of the global economy and a climate system that includes GHG emissions, carbon dioxide concentration, climate change, climate change impact and optimal policy. 
Compared to DICE, the most notable feature of RICE is dividing the world into multiple regions.

IAMs are often criticized for having limited economic relationship between regions, failing to adequately account for each country's industrial balance, failing to incorporate endogenous technological advancement, and treating uncertainty deterministically. 
Furthermore, most IAMs struggle to incorporate the scale of the scientific risks as temperature rises (non-linearity), like the thawing of permafrost, release of methane, or any other potential tipping points.

In one of the most recent examples, the United States government~\cite{InteragencyWorkingGrouponSocialCostofGreenhouseGases2021} produced SCC estimates based on three different IAMs: DICE\footnote{Dynamic Integrated Climate and Economy\cite{Nordhaus2020}}, PAGE\footnote{Policy Analysis of the Greenhouse Gas Effect\cite{Hope2011}} and FUND\footnote{Climate Framework for Uncertainty, Negotiation, and Distribution\cite{Tol}}.
These three IAMs, built in a near similar fashion, involve four modules — socioeconomic, climate, damages, and discounting — to estimate the impact of climate change on civilization.

These IAMs, however, differ in two significant ways: how they estimate impact of emissions on global temperature \cite{EPRI2014a} and how they estimate the expected damages from climate change~\cite{Moore2015a}. 
The damage module maps changes in climate to economic damages.
While this module is central to SCC estimates: it is often the most simplified element as well, which can cause systematic bias in the SCC estimates produced by these models. 
Therefore, even with the same initial conditions, annual global damages across each model could substantially differ. 
Additionally, these IAMs are extremely sensitive to their initial conditions, suggesting that small errors in the initial calibration can have a large impact of the predicted damages\cite{platt_comparison_2019}.

Users typically take the difference of these models to guide decisions. 
While some \cite{NationalAcademiesofSciences2017} argue such differences are more of “a result of uncoordinated modeling choices,” the reliance on diverging models places the use of IAMs into question. 
IAMs have come under increasing criticism \cite{Revesz2014GlobalChange} due to their lack of transparent inputs \cite{Pfenninger2017} and inability to account for irreducible uncertainty \cite{ClimateChangePolicy}, among other issues \cite{Farmer2015}. 

To address IAMs’ shortcomings, and implications for the climate and economy, some critics \cite{Weyant2017} suggest calling for increased transparency, some \cite{ClimateChangePolicy} propose that expert judgement should take more precedence, and some \cite{Farmer2015} point to new modeling approaches. 

Recently, newer IAMs have made progress in addressing these criticisms. 
MRICES~\cite{wang2016ciecia, wang2010policy} has introduced the linear mechanism of Mundell-Fleming, yet it does not fully reflect the complexity of the economic inter linkages among different regions of the world. 
The Capital, Industrial Evolution and Climate change Integrated Assessment model (CIECIA)~\cite{wang2016ciecia} is one of the first to develop a balanced economy IAM. 
CIECIA divides the entire world into several countries (or group of country), and in every country, the economic system is decomposed in 12 sectors. 
~\citet{cai2013social} show the importance of tackling uncertainty. 
They showed that models without uncertainty tend to significantly underestimate the benefits of abatement policies. 
IAMs deal with uncertainty in a probabilistic manner, by means of Monte Carlo analysis~\cite{nordhaus2007challenge, kypreos2007merge, ackerman2010fat}. 
That said, their simulation approach does not reflect the impact of uncertainty on decision making.

The economic impacts of climate change differ considerably depending on model assumptions. 
For example, models that include market frictions and/or agents that make imperfect decisions due to information asymmetries tend to exhibit greater variation in outcomes than models that assume that agents are rational, welfare-maximising and have perfect foresight. 
Recent works by the Network for Greening the Financial System (NGFS, 2020) estimates that the transition to a low carbon economy could reduce world economic activity by about 4 percent in orderly scenarios (immediate actions) to slightly less than 10 percent in disorderly ones (delayed responses).

\subsection{Market-based approaches}

There has been emerging initiatives in forming market- or incentive-based approaches in addressing climate change concerns. 
Market-based innovations in derivatives markets and climate finance options are vital, bringing together public and private sectors to work together in creating synergy for the long-run benefit of the society. 
It is, therefore, crucial for policy makers and scientists to develop new tools that leverage multidisciplinary expertise in innovating tools to advance this effort. 
In this review, we have focused on two key streams: (i) cap-and-trade and some key markets; and (ii) voluntary carbon markets. 

Emission trading, sometimes referred to as ``allowance trading" or ``cap-and-trade", is a market-based approach that consist of two key elements: (i) a limit (or cap) on pollution, and (ii) tradable allowances equal to the limit that authorize permit holders to emit a specific amount of the pollution. 
The former ensures that the environmental goals can be met. 
The latter provides sufficient flexibility for individual emission sources (plants or firms) to set their own compliance path. 
Because permits can be traded in a market and the market price is determined by the supply and demand at equilibrium, cap-and-trade programs are often referred as the early forms of market-based instruments.

There exist several key global cap-and-trade markets. 
After over a decade of research and extensive public debate since 1980, the US Congress established the Acid Rain Program under its 1990 Clean Air Act Amendments. 
The program called for major reduction in electric-generating facilities' emissions of sulfur dioxide (SO$_2$) and nitrogen oxides (NO$_x$) the key components of acid rain. 
This program sets a national cap on SO$_2$ emissions, while authorizing emission allowance based on emission performance and representative fuel use. 
Each plan will be able to choose many alternatives ranging from deploying energy-efficiency measure and install pollution control equipment, or purchasing excess allowance from other sources. 

Under cap-and-trade program, all trades under the cap represents a reduction in total emission. 
Because of the flexibility in the options firms can choose to reduce emission, this design allows them to reduce emission at the lowest possible cost. 
In fact, trading is generally a small component of a firm's overall strategy for meeting emission limits as evidence has shown that largest polluters tend to significantly reduce their emission before purchasing allowance from other firms. 
In addition to reducing the overall amount of emission, cap-and-trade also accelerate the pace of emission reduction plan of firms. 
This benefit can be attributed to the marginal benefit of emission reduction that is associated with a monetary value through trading.
Firms have greater incentives in reducing emission sooner than required in order to avoid the increasing future costs of purchasing a permit at a higher price down the road. 
This also result in earlier realization of environment benefits. 

Similar to the early market-based climate innovation through cap-and-trade, The EU Emissions Trading System (ETS) is the cornerstone of the EU policy to combat climate change. 
ETS is designed to reduce emissions with minimal cost to society, while stimulating technological innovation to further mitigate this cost in the future. 
The EU ETS was designed in three phases, with policies governing the operation of each phase informed by the one that came before.
Phase 1 came into effect in 2005 as a three-year pilot period. 
Phase 2 ran from 2008-2012 and saw an expansion in coverage of both countries and sectors. 
Currently in Phase 3 more than 12,000 power and industrial plants in 31 countries are taking part in the scheme. 
While the third phase was scheduled to end in 2020, some policies were extended beyond the end of the scheme. 
At the beginning of Phase 3, the upper limit on total emissions was set to decline at a rate of 1.74\% per year up until 2020 and 2.2\% per year until 2030. 
By this time, EU emissions will be 43\% less than they were in 2005. 

Participating firms of EU-ETS are required to surrender one pollution permit, known as an EU allowance (EUA), for each metric ton of CO2 (tCO2e) emitted. 
These permits are distributed to companies either for free or through an auctioning system. 
The market-based nature of the EU-ETS is reflected through the supply and demand equilibrium. 
The total maximum number of permits available – the ‘cap’ – is designed to limit emissions below the levels that would otherwise be produced – this is referred to as the ‘business-as-usual’ emissions. 
This would be the emission level when the economy is operating at its production potential. 
The ETS has also established an EU-wide carbon price that signals the opportunity-cost of emitting CO2 to all carbon market participants. 
By design, a scarcity of permits (short of supply) will drive up their price, which companies buy or sell on the market. 
There is also an incentive for participants to reduce their emissions up to the indifference point – a level at which there is no difference between buying one permit at the current market price and paying the cost of reducing emissions by one additional ton of carbon. 
As each market participants evaluates its own projected emission (through forecasting production and investment), the optimal choices of demand for permit will determine a ``fair" market price for the permit, indirectly the price of carbon emission. 
While the ``cap" tries to reflect some long-run cost of emission, the cap-and-trade program can't fully resolve the Tragedy of Horizon problem due to key limitations such as imperfect technology (of estimating the carbon costs) and the imperfect information (among participating firms).

More recently, large, high-integrity, voluntary carbon markets (VCM) play an important role in mitigating climate change by enabling billions of dollars flowing to projects that help avoid, reduce, remove and sequester CO2 emissions. For example, the Taskforce on Scaling Voluntary Carbon Markets (TSVCM) is a private sector-led initiative working to scale an effective and efficient voluntary carbon market to help meet the goals of the Paris Agreement.
In order to accelerate climate action, we need investments in both avoidance and reduction, as well as removal and sequestration credits. 
TSVCM supports three dimensions of the policy aspect of climate change, notably: 

\begin{itemize}
    \item Governance: through the establishment of an umbrella governance body to oversee the VCMs landscape, to increase credit quality and harmonize standards, and
    \item Legal \& contracts: to define standardized reference terms of the core carbon that drive liquidity in the market, thereby supporting a transparent price signal, and 
    \item Credit-level Integrity:  to define Core Carbon Principles and additional attributes that ensure high quality standards and credibility. 
\end{itemize}

VCM advocates for voluntary disclosure and participation so that market reveals information from participating institutions and corporations on emission level and associated short- and long-run cost.
Traditional climate policy doesn't address the incomplete information problem where the lack of full market information could precisely reflect the combined assets value of each agent (corporate firm and household). 
Under VCM, households (creditors) would be able to supply funds towards more green corporations, this increase incentives for green firms to further enhance technology to ensure the emission-adjusted stock value reflects a competitive return.
This further improves the balance sheet of corporations, therefore improving the balance sheet of individual household. 
Taking the economy as a whole, this market is the most efficient in limiting the climate risk that the economy and more importantly, financial markets of different country will face.  
To the best of our knowledge, there lacks theoretical framework and simulations evidence to address this contracts objective of VCM. 
We hope more efficient utilization of AI/ML tools in climate change can help abstract a market-based price signal thereby improving the design of climate policy.  

\section{Aleatoric and Epistemic Uncertainty : an Example Using Linear Regression}

Consider the following stochastic relation:

$$ y_i = \mathbf{x}_i^T \pmb{\beta} + \epsilon_i, \qquad \mathbf{x_i} \in \mathcal{X}, y_i \in \mathcal{Y}$$

We suppose that we do not know \textit{parameters} $\pmb{\beta}$ : the relation is \textit{unobserved}.
When trying to estimate $\pmb{\beta}$, a typical approach in economics involves collecting a set of samples $ \{(\mathbf{x_i},y_i)\}_{i=1}^n$ and fitting a linear regression model to these samples:

$$ \hat{y}_i = \mathbf{x}_i^T \mathbf{b} + e_i $$

In other words, we want $\mathbf{b}$ to be as similar as possible to $ \beta$.

However, our present interest lies more in the latter parts of both equations.
The term $\epsilon_i$ is called the \textit{disturbance}, whereas $e_i$ is the \textit{residual}.

The disturbance represents the randomness present in the stochastic relation itself.
Such randomness cannot be removed, no matter how close our estimate of $b$ is to $\beta$.
This kind of randomness, when considered across all possible data points, is referred to as \textit{aleatoric} (or aleatory) uncertainty.
It is also commonly referred to as the irreducible error.

The residual $e_i$ is a combination of the disturbance due to aleatoric uncertainty and an additional term due to the \textit{epistemic} uncertainty, which is any type of uncertainty that is reducible by increasing information.
For example, the estimate $b$ of $\beta$ could be improved by collecting additional data points or by using better estimation methods, which would lower the residual across all possible data points.
As epistemic uncertainty decreases, the average residual approaches the aleatoric uncertainty.





\section{Toy Experiment: ReSCCU}

\begin{figure}[!ht]
     \centering
     \begin{subfigure}[b]{0.48\textwidth}
         \centering
         \includegraphics[width=\textwidth]{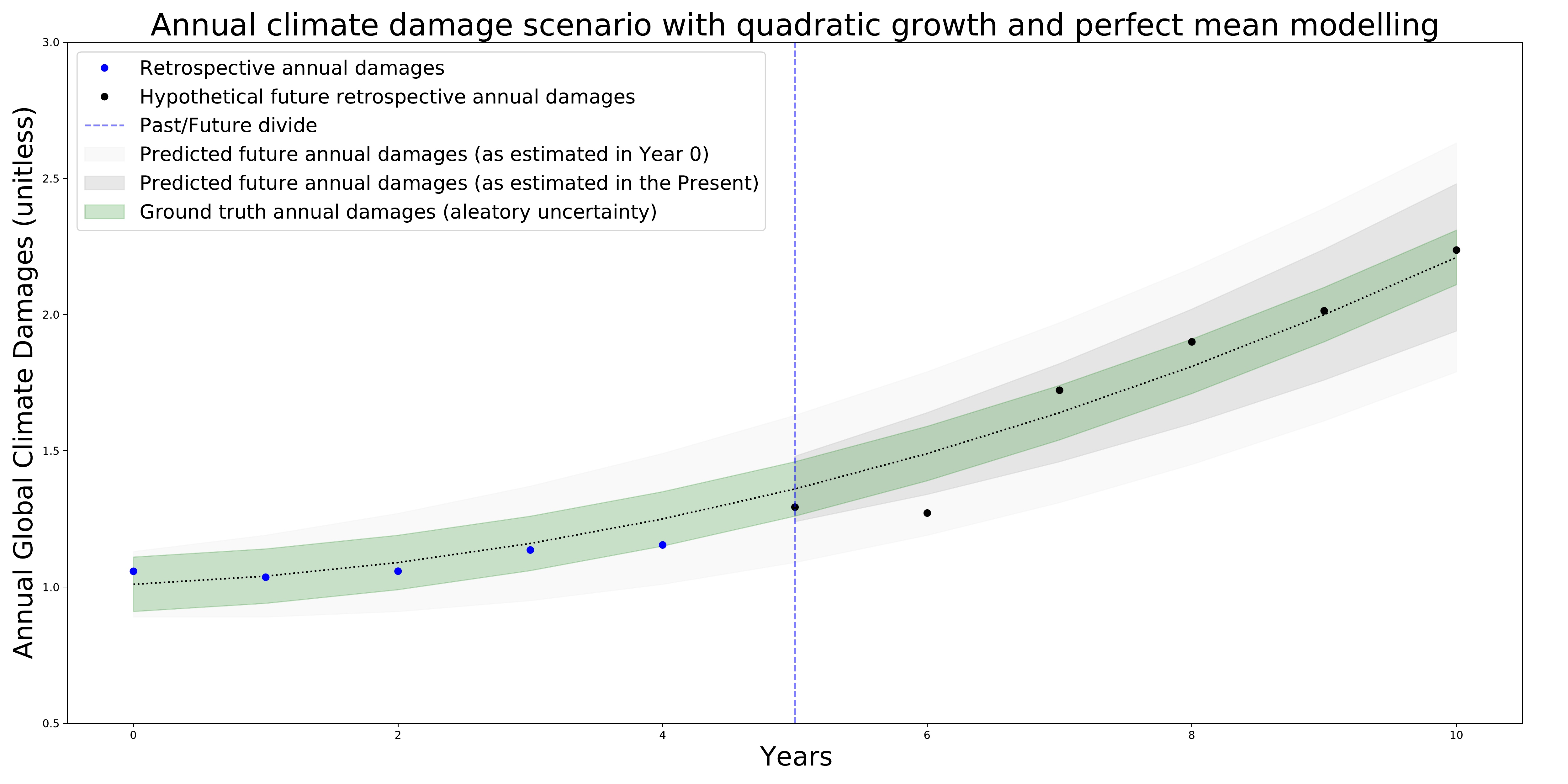}
         \caption{A ReSCCU scenario under the assumption that the SCC model predicts the damage perfectly in the mean.}
         \label{fig:y equals x}
     \end{subfigure}
     \hfill
    \begin{subfigure}[b]{0.48\textwidth}
         \centering
         \includegraphics[width=\textwidth]{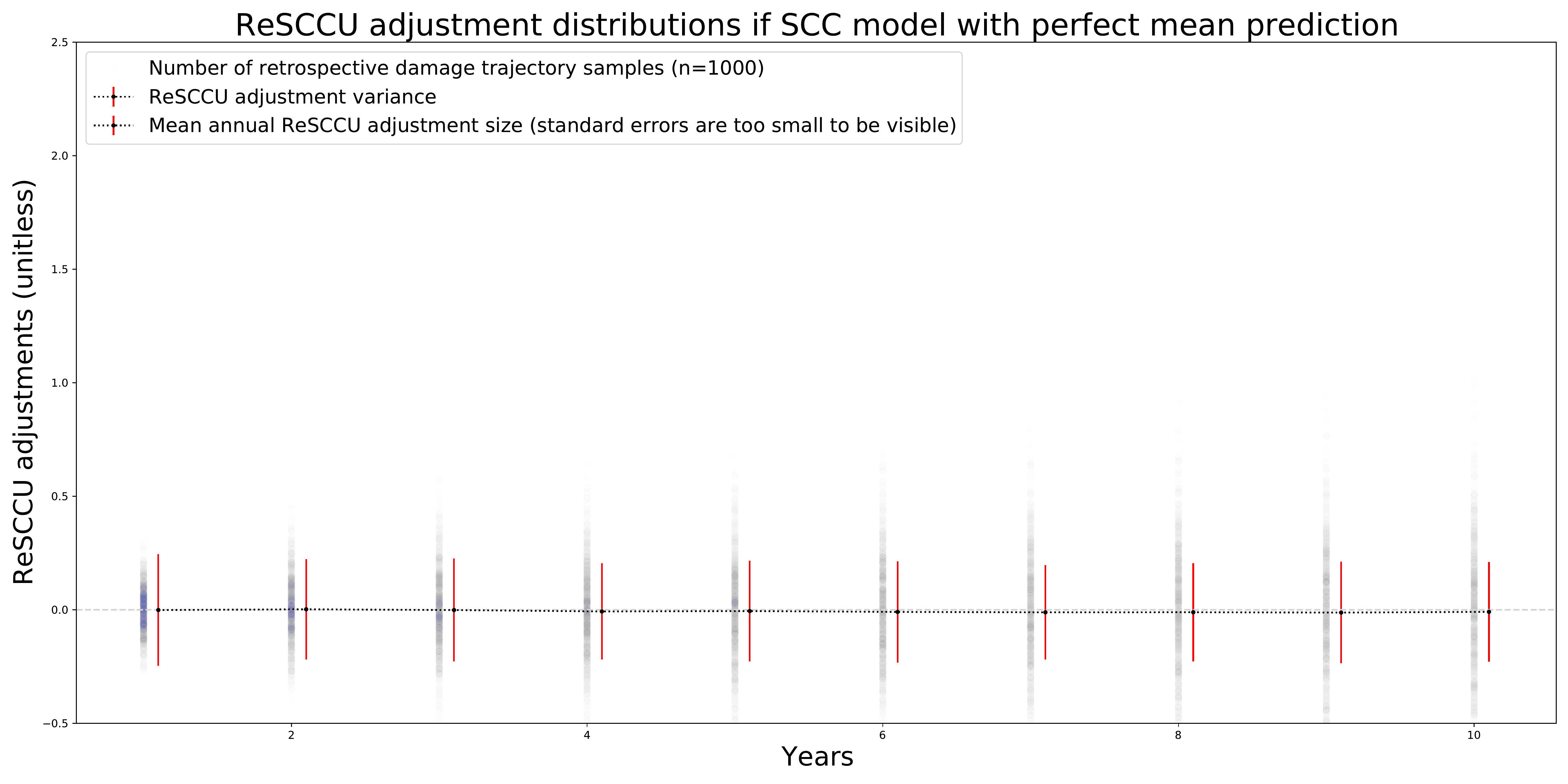}
         \caption{Average ReSCCU readjustments under the assumption that the SCC model predicts the damage perfectly in the mean.}
         \label{fig:y equals x 2}
     \end{subfigure}
     \medskip
     \begin{subfigure}[b]{0.48\textwidth}
         \centering
         \includegraphics[width=\textwidth]{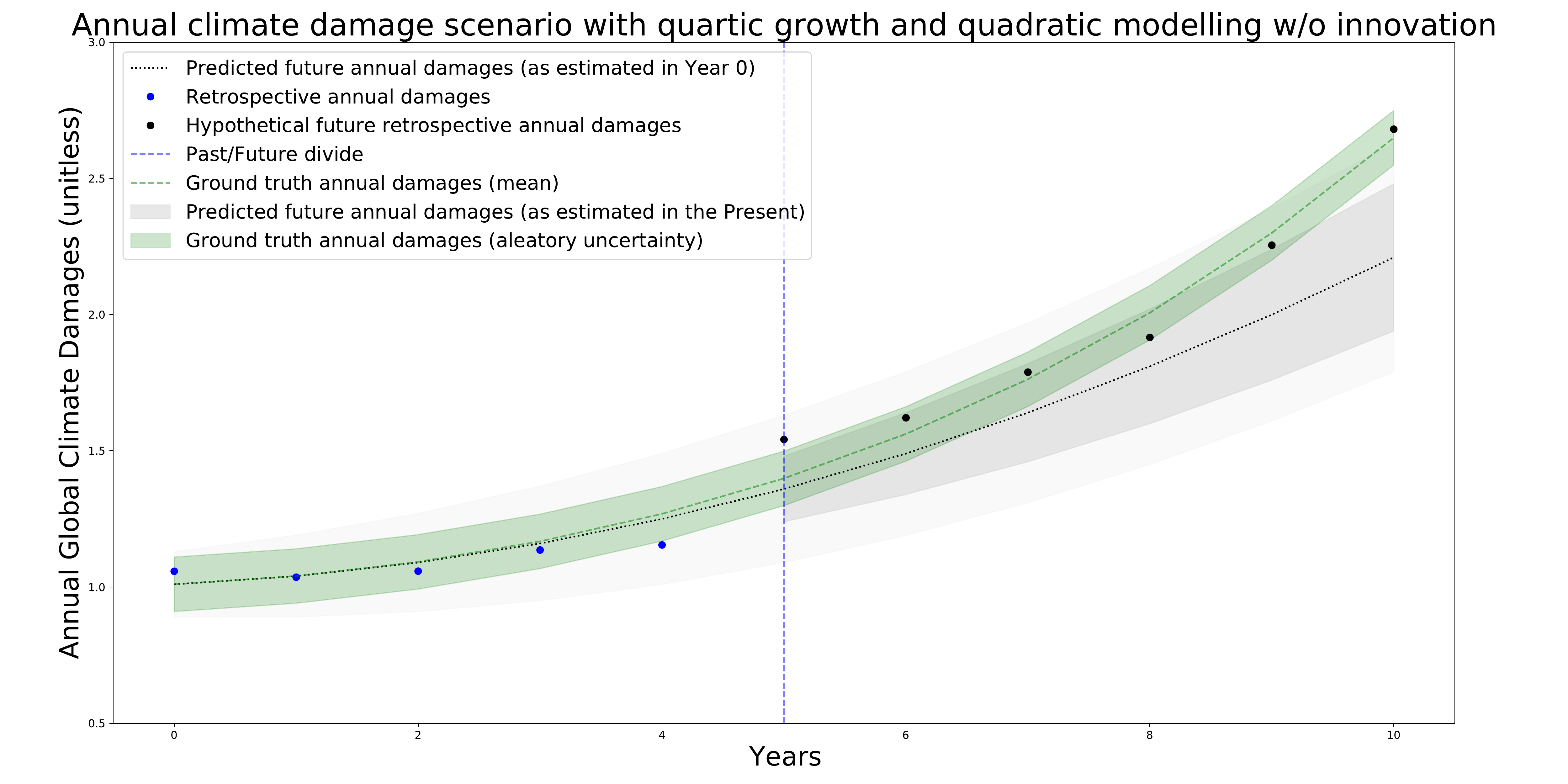}
         \caption{A ReSCCU scenario under the assumption that the SCC model fails to capture a quartic non-linearity.}
         \label{fig:three sin x}
     \end{subfigure}
     \hfill
     \begin{subfigure}[b]{0.48\textwidth}
         \centering
         \includegraphics[width=\textwidth]{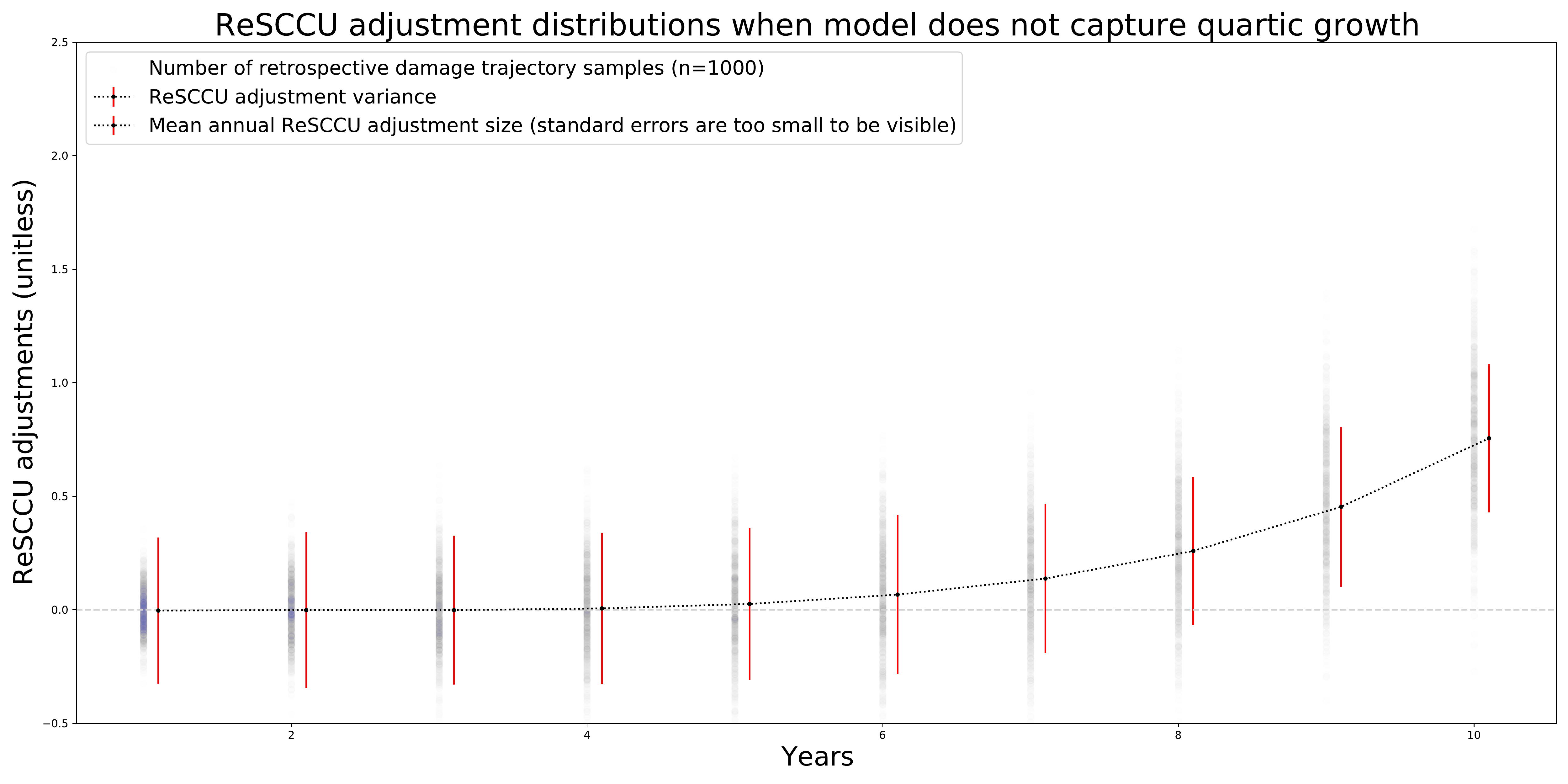}
         \caption{Average ReSCCU readjustments under the assumption that the SCC model fails to capture a quartic non-linearity.}
         \label{fig:three sin x 2}
     \end{subfigure} 
      \medskip
     \begin{subfigure}[b]{0.48\textwidth}
         \centering
         \includegraphics[width=\textwidth]{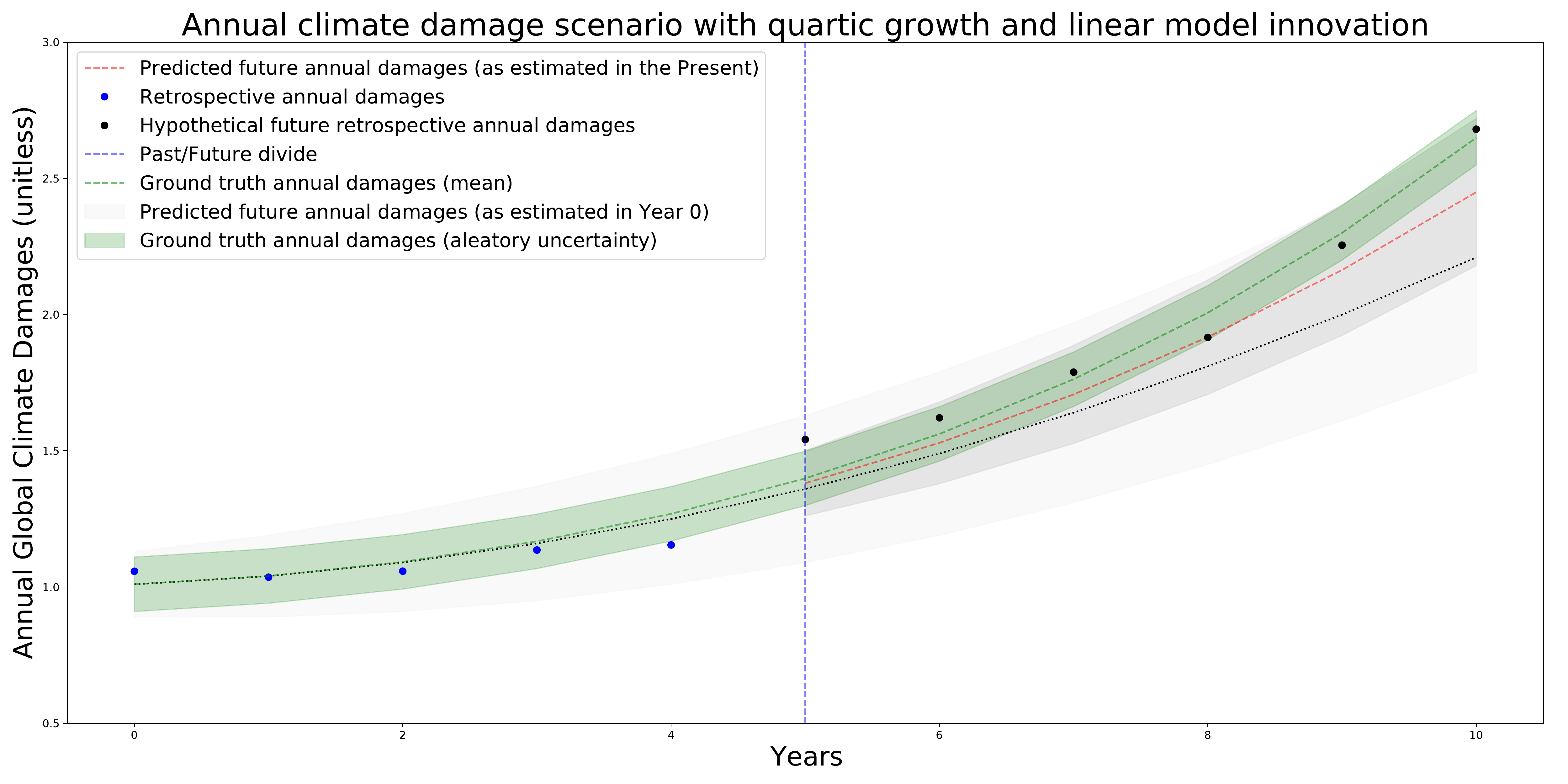}
         \caption{A ReSCCU scenario under the assumption that the SCC model innovates toward a quartic non-linearity.}
         \label{fig:five over x}
     \end{subfigure}
     \hfill
     \begin{subfigure}[b]{0.48\textwidth}
         \centering
         \includegraphics[width=\textwidth]{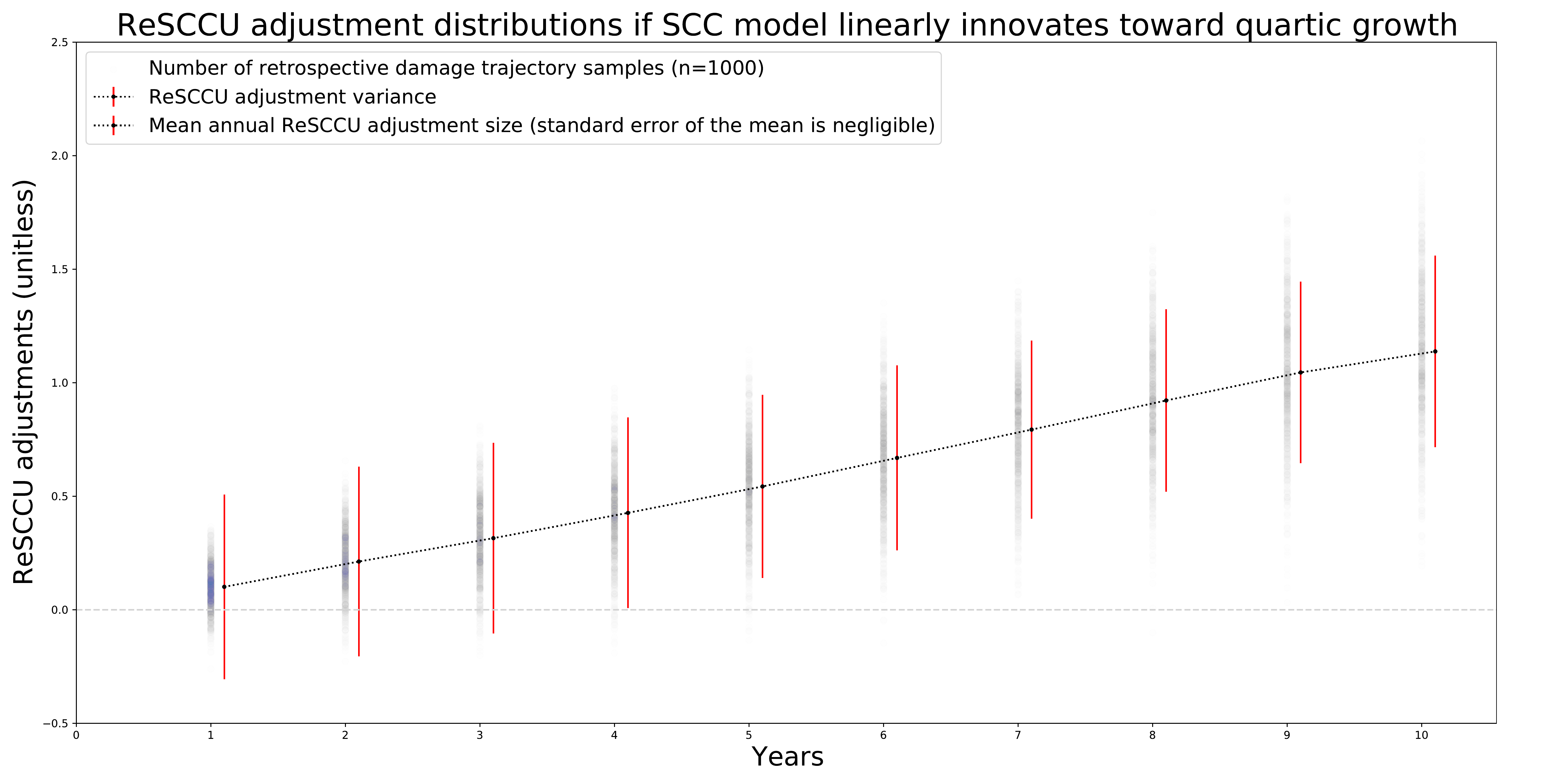}
         \caption{Average ReSCCU readjustments under the assumption that the SCC model innovates toward a quartic non-linearity.}
         \label{fig:five over x 2}
     \end{subfigure}
        \caption{An analysis of three abstract scenarios in which SCC models are able to predict real-world damages to different extents, and the associated ReSSCU adjustments.}
       \label{fig:recap}
\end{figure}

Figure \ref{fig:recap} provides a graphical illustration of ReSCCU. 

We see from a) and b) that average retroactive adjustments $\mathbb{E} \left[\Delta c_t\right],\ \forall t$ are zero if we assume that our SCC prediction model is perfect in the mean. However, $\mathbb{V}ar  \left[\Delta c_t  \right] \neq 0$, emphasising the point that SCC estimates will almost surely not equal cumulative future damages if time horizons are finite and the environment carries aleatory uncertainty. 
Subfigures c)-f) then assume that the (originally quadratic) damage function has a slow-onset quartic component that is initially not represented in the SCC prediction function.
If no model innovation happens over time, then d) illustrates that the size of SCC adjustments will grow increasingly fast with time, motivating the need for the ReSCCU framework. 
In reality, one might assume that the SCC prediction model may be innovated to reflect the quartic component as it is started to be observed in retrospective damage measurements. 
Assuming that this model innovation process evolves as a linearly annealing mixture between the true and the initial prediction models, f) demonstrates that ReSCCU adjustments will grow faster initially but remain at constant growth, illustrating how (constructive) SCC model innovation helps correct epistemic modeling errors earlier but does not obviate the need for ReSCCU adjustments. \\

We can see that, over time, ReSCCU estimates become increasingly grounded in retrospective marginal damages, meaning that estimated climate damages are gradually replaced by damages actually caused. 
If we presume a finite $T$, then approaching $t=T$ and within the limitations of climate damage measurements, the ReSCCU converges to the true SCC. 
Note, however, that the rate and monotonicity of this convergence is constrained by the characteristics of the future marginal damage estimation functions.
 
 \section{PReCaP: Derivation of the SCC signal} 

The RetroExchange observes the insurance premium spread at each transaction (independently of whether UICC to ICC conversion takes place or not).
The insurance premium spread itself is, however, confounded by

\begin{centering}
\begin{enumerate}
    \item speculators or bets on the future SCC
    \item finite horizon of the NI
    \item price adjustments aiming to hedge against future business risks, or recovering from past mistakes. 
\end{enumerate}
\end{centering}

Past mistakes can be detected by the RetroExchange (and hence potentially adjusted for), and insurance hedging behavior should be minimized by the competition between insurers.
In this way, it is ensured that 

\begin{centering}
\begin{enumerate}
\item each carbon credit traded on the RetroExchange is insured exactly once and 
\item both buyers and sellers are completely unexposed to the retroactive carbon pricing mechanisms - to them, the RetroExchange feels like an ordinary exchange.
\end{enumerate} 
\end{centering}

Carbon credits can be traded across other exchanges as long as any UICC is first converted to an ICC before traded on the RetroExchange (ICCs can be sold at any other exchange if needed - to other exchanges, they are just ordinary carbon credits).

\end{document}